\documentclass[11pt,a4paper]{article}
\usepackage{jheppub, braket, tikz, color, multirow, subfigure}
\usepackage[all]{xy}

\def\K{{\cal K}}
\def\Q{{\rm Q}}
\def\L{{\cal L}}
\def\P{{\cal P}}

\def\H{{\cal H}}
\def\V{{\cal V}}
\def\M{{\cal M}}

\def\1{{\bf 1}}
\def\Z{{\mathbb{Z}}}
\def\N{{\cal N}}
\def\psl{\mathfrak{psl}(2|2)}
\def\g{\mathfrak{g}}
\def\bg{\mathfrak{g}^{(0)}}
\def\id{{\mathrm{id}}}
\def\qt{{\mathcal{Q}}}

\newcommand{\ol}[1]{\overline{#1}}
\newcommand{\mycaption}[1]{\caption{\textsl{#1}}}
\newcommand\rep[3]{\L_{#3}\bigl(#1\bigr)}
\newcommand\tp[2]{{#1 \otimes \ol{#2}}}

\title{The massless string spectrum on AdS$_3\times {\rm S}^3$ from the supergroup}

\author{Matthias R.\ Gaberdiel}
\author{and Sebastian Gerigk}

\affiliation{
Institut f\"ur Theoretische Physik, ETH Zurich, \\
CH-8093 Z\"urich, Switzerland}

\emailAdd{gaberdiel@itp.phys.ethz.ch}
\emailAdd{gerigk@itp.phys.ethz.ch} 

\abstract{String theory on AdS$_3\times {\rm S}^3$ is studied in the hybrid formulation. We give 
a detailed description of the PSL(2$|$2)  supergroup WZW model that underlies the
background with pure NS-NS flux, and determine the BRST-cohomology corresponding to 
the massless string states. The resulting spectrum is shown to match exactly with the expected 
supergravity answer, including the sectors with small KK momentum on the sphere.}

\begin{document}

\maketitle

\section{Introduction}

One of the simplest examples of the AdS/CFT correspondence \cite{Maldacena:1997re}
is the duality between superstring theory on AdS$_3\times {\rm S}^3$, and a 2-dimensional
conformal field theory living on the boundary of AdS$_3$. Many properties of this
duality can be studied  in quite some detail since both sides of the 
correspondence are under very good control. This is, in particular, the case if the AdS
background has pure NS-NS flux, since the corresponding world-sheet theory 
has an NS-R formulation in terms of a WZW model 
\cite{Giveon:1998ns, Kutasov:1999xu, Kutasov:1998zh} whose structure has been studied in quite
some detail \cite{Maldacena:2000hw,Maldacena:2000kv,Maldacena:2001km}. Using this
approach, many detailed checks of the correspondence have been performed, for example,
three-point functions have been compared 
\cite{Gaberdiel:2007vu,Dabholkar:2007ey,Pakman:2007hn,Taylor:2007hs,Cardona:2009hk,Giribet:2007wp,
deBoer:2008ss}.

The main drawback with this approach, however, is that it is difficult to switch on R-R flux. 
In order to overcome this limitation, the hybrid formulation  \cite{Berkovits:1999im} was 
developed (see also \cite{Berkovits:1999du,Berkovits:1999xv}), in which the 6d 
AdS$_3\times {\rm S}^3$ part of the background is described in a 
Green-Schwarz-like  formulation, while the remaining 4d background is treated using NS-R variables. 
More specifically, the internal AdS$_3\times {\rm S}^3$ background can then be formulated in terms of 
a sigma-model on the supergroup PSL$(2|2)$ for which spacetime supersymmetry is manifest
\cite{Metsaev:1998it, Rahmfeld:1998zn,Bershadsky:1999hk}. 
Non-linear sigma-models with supergroup targets and their cosets have attracted a lot of attention
recently, and some of their properties have been studied 
\cite{Rozansky:1992rx,Read:2001pz,Gotz:2005ka,Schomerus:2005bf,Gotz:2006qp,Saleur:2006tf}. 
However, apart from 
the recent analysis of  \cite{Troost:2011fd}, the relevant techniques have not yet been employed in
the hybrid formulation of AdS$_3\times {\rm S}^3$. It is the aim of the present paper
to make progress along these lines. In particular, we shall give a detailed description 
of the supergroup WZW model (that  corresponds to the background with pure NS-NS flux),
and check that the massless string spectrum it describes matches exactly the 
expected supergravity answer.
\smallskip

For the case with pure NS-NS flux, the supergroup sigma-model can be described in terms of 
a (supergroup) WZW model that defines a logarithmic conformal field theory (LCFT) 
\cite{Schomerus:2005bf,Gotz:2006qp,Saleur:2006tf}. Using ideas that had been developed before for the 
analysis of the logarithmic triplet models in \cite{Gaberdiel:2007jv,Gaberdiel:2009ug},
we make a detailed proposal for the spectrum of this LCFT, extending the analysis of 
\cite{Schomerus:2005bf,Gotz:2006qp,Saleur:2006tf,Troost:2011fd}. The key step involves determining the 
projective covers for the representations of interest, and while most of this analysis proceeds as 
in the finite dimensional case following \cite{Zou:1996,Germoni:1998} (see also 
\cite{Gotz:2006qp,Troost:2011fd}),  there are some important differences for the projective covers 
of small momenta that we shall explain in detail. Once the structure of the projective covers is under
control, there is a natural proposal for how the left- and right-moving projective 
representations have to be coupled together, leading to a description of the full spectrum
as the quotient space of the direct sum of  tensor products of the projective representations. 
This fixes the spectrum of the underlying world-sheet CFT, from which one can then 
obtain the string spectrum as a suitable BRST cohomology. 

In order to check our proposal we then calculate the BRST cohomology for the massless string states. 
For this case the BRST cohomology was previously studied in terms of vertex operators in 
\cite{Dolan:1999dc}. We explain how the BRST operators of  \cite{Berkovits:1999im,Dolan:1999dc} 
can be lifted to act on the projective covers (from which the LCFT spectrum can be obtained by 
quotienting). It is then straightforward to determine their common cohomology, and hence the 
massless physical string spectrum. We find that the resulting spectrum agrees precisely with the 
supergravity prediction of  \cite{Deger:1998nm,de Boer:1998ip}, including the truncations that appear for small 
momenta. We should mention that the same problem was also recently attacked in  
\cite{Troost:2011fd}, where, however, the analysis was only performed for sufficiently large 
momenta, and the precise way in which left- and right-moving representations are coupled 
together was only sketched. 
\smallskip

The paper is organised as follows. We start in Section~\ref{rep_psl} by reviewing the basics of 
Lie superalgebras \cite{Kac:1977em} and their representations. We explain the structure of the
irreducible and the Kac modules \cite{Kac:1977hp} of interest, and then make a proposal
for the corresponding projective covers. 
In Section~\ref{physical_states} we  explain how to construct the full space of states of the 
logarithmic conformal field theory of the WZW model based on PSL$(2|2)$, following recent
ideas of  \cite{Gaberdiel:2007jv,Gaberdiel:2009ug}. We then explain how the BRST operator on 
massless string states  
\cite{Berkovits:1999im, Dolan:1999dc} can be formulated in our language, and study its
cohomology. Finally, we show that this BRST cohomology reproduces precisely the physical
spectrum of  ${\cal N}=2$ supergravity in six dimensions \cite{Deger:1998nm,de Boer:1998ip}. Section~\ref{concl}
contains our conclusions, and our conventions for the description of the superalgebra
are spelled out in the  Appendix.

\section{Representations of $\psl$} \label{rep_psl}

Let us begin by reviewing the representation theory of $\g = \psl$; this will
also allow us to fix our notations.

\subsection{The Lie Superalgebra} \label{intro_superalg}

Like any Lie superalgebra, $\psl$ allows for a decomposition into bosonic and fermionic generators 
$\g = \bg \oplus \g^{(1)}$ \cite{Scheunert:1979xy}, where $\bg$ is the bosonic Lie subalgebra 
$\bg = \mathfrak{sl}(2) \oplus \mathfrak{sl}(2)$. Furthermore, $\psl$ is a Lie superalgebra of type I, 
which means that the fermionic summand $\g^{(1)}$ can be further decomposed as 
$\g^{(1)} = \g_{-1} \oplus \g_{1}$ such that
\begin{equation}
\{\g_{-1},\g_1\} \subset \bg \ , \qquad \{\g_1,\g_1\} = \{\g_{-1},\g_{-1}\} = 0 \ .
\end{equation}
This decomposition introduces a natural grading $\rho$, where $\rho(\g_{\pm1}) = \pm 1$ and 
$\rho(\bg) = 0$. $\rho$ lifts to a $\Z$-grading on the universal enveloping algebra ${\cal U}(\g)$ in the 
obvious way. An explicit description of the generators and their commutation relations is given in 
Appendix~\ref{app:A}.

\subsection{Kac Modules and Irreducible Representations}

For comparison to string theory on AdS$_3 \times {\rm S}^3$ we will mainly be interested in 
representations whose decomposition which respect to the bosonic subalgebra 
$\bg = \mathfrak{sl}(2) \oplus \mathfrak{sl}(2)$ leads to infinite-dimensional discrete series 
representations with respect to the first $\mathfrak{sl}(2)$ (that describes isometries
on ${\rm AdS}_3$),  and finite-dimensional representations with respect to the second 
$\mathfrak{sl}(2)$ (that describes isometries of ${\rm S}^3$).
As in \cite{Gotz:2006qp} we shall label them by a doublet of half-integers 
$(j_1, j_2)$ where $j_1 \leq - \tfrac{1}{2}$ and $j_2 \geq 0$. The cyclic state of the corresponding
representation is then characterised by 
\begin{equation}
\begin{array}{c}
J^0 \ket{j_1,j_2} = j_1 \ket{j_1, j_2} \ , \qquad K^0 \ket{j_1,j_2} = j_2 \ket{j_1, j_2} \ , \\[0.25cm]
J^+ \ket{j_1,j_2} = K^+ \ket{j_1, j_2}=\left(K^-\right)^{(2j_2 + 1)}  \ket{j_1, j_2} = 0 \ .
\end{array}
\end{equation}
Here $J^0,J^\pm$ are the generators of the first $\mathfrak{sl}(2)$ with commutation relations 
\begin{equation}
{}[J^0,J^\pm] = \pm J^\pm \ , \qquad [J^+,J^-] = 2 J^0 \ ,
\end{equation}
while $K^0,K^\pm$ are the generators of the second $\mathfrak{sl}(2)$ that satisfy identical commutation 
relations. We denote the corresponding highest weight representation of 
$\bg = \mathfrak{sl}(2) \oplus \mathfrak{sl}(2)$ by $\V(j_1,j_2)$. 

Each representation $\V(j_1,j_2)$ of $\bg$ gives rise to a  representation of the full Lie superalgebra
by taking all the modes in $\g_{+1}$ to act trivially on all states in $\V(j_1,j_2)$, $\g_{+1} \V(j_1,j_2)=0$, and 
by taking the modes in $\g_{-1}$ to be the fermionic creation operators. The resulting representation is usually 
called the Kac module  \cite{Kac:1977hp} and will be denoted by $\K(j_1,j_2)$. The dual construction, where
$\g_{+1}$ are taken to be the fermionic creation operators while $\g_{-1}$ are annihilation operators, defines
the dual Kac module $\K^\vee(j_1,j_2)$. 
The grading $\rho$ of the universal enveloping algebra ${\cal U}(\g)$ induces a grading on the Kac module, 
where we take all states in $\V(j_1,j_2)$ to have the same grade, say $g\in \Z$.  If we want to stress this 
grade assignment, we shall sometimes write $\K_g(j_1,j_2)$. The states involving one fermionic generator 
from $\g_{-1}$ applied to the states in $\V(j_1,j_2)$  then have grade $g-1$, {\it etc}.
\smallskip

The Kac modules  are either typical or atypcial \cite{Kac:1977hp}. We call the Kac module $\K(j_1,j_2)$ 
typical if it is irreducible. This is the case for generic values of $j_1$ and $j_2$, and then the
corresponding irreducible representation $\L(j_1,j_2)$ is simply equal to $\L(j_1,j_2)=\K(j_1,j_2)$.
On the other hand, if $\K(j_1,j_2)$ is reducible, the Kac module is called atypical. In the case at hand, 
{\it i.e.}\ for $j_1<0$ and $j_2\geq 0$,  the Kac module  $\K(j_1,j_2)$ is atypical if and only if 
\cite{Gotz:2006qp}
\begin{equation}
j_1 + j_2 + 1 = 0 \ .
\end{equation}
This condition is equivalent to the condition that the quadratic Casimir $C_2$ vanishes on the Kac module.
We shall denote atypical Kac modules by a single index, $\K(j)\equiv \K(-j-1,j)$. The corresponding irreducible 
representation $\L(j)\equiv \L(-j-1,j)$ is then the quotient of $\K(j)$, where we devide out the 
largest proper subrepresentation $M_1$ of $\K(j)$
\begin{equation}
\L(j) = \K(j)/ M_1   \ .
\end{equation}
In the following we shall almost exclusively  consider the atypical representations, since these are the 
only representations that matter for the massless string states. The structure of the corresponding 
irreducible representations (with respect to the action of the bosonic
subalgebra $\g^{(0)}$) is described in Fig.~\ref{irr_g0}. 
\begin{figure}[h]
\begin{center}

\begin{tikzpicture}

\node (name) at (-5,1) {$\L_0(j \geq \frac{1}{2}):$};

\node (top) at (0,2) {$(-j-1,j)_0$};
\node (left) at (-2,1) {$(-j-\frac{3}{2}, j-\frac{1}{2})_{-1}$};
\node (right) at (2,1) {$(-j-\frac{1}{2},j+\frac{1}{2})_{-1}$};
\node (bottom) at (0,0) {$(-j-1,j)_{-2}$};

\draw (top) to (left);
\draw (top) to (right);
\draw (bottom) to (left);
\draw (bottom) to (right);

\node (name) at (5,1) {$\L_0(0):$};

\node (top) at (7,2) {$(-1,0)_0$};
\node (right) at (8,1) {$(-\frac{1}{2}, \frac{1}{2})_{-1}$};
\node (bottom) at (7,0) {$(-1,0)_{-2}$};

\draw (top) to (right);
\draw (bottom) to (right);
\end{tikzpicture}

\mycaption{The decompostion of atypical irreducible $\g$-representations into $\g^{(0)}$-components.}
\label{irr_g0}

\end{center}
\end{figure}

The atypical Kac modules are reducible but not completely reducible. In order to describe their structure it 
is useful to introduce their composition series. This keeps track of how the various subrepresentations sit
inside one another. More precisely, we first identify the largest proper subrepresentation $M_1$ of $\K(j)$, 
so that $\L(j) = \K(j)/ M_1$ is irreducible; we call the irreducible representation $\L(j)$  the {\em head} of 
$\K(j)$.  Then we repeat the same analysis with $M_1$ in place of $\K(j)$, {\it i.e.}\ we identify the 
largest  subrepresentation $M_2$ of $M_1$ such that $M_1/M_2$ is a direct 
sum of irreducible representations. The composition series is then simply the sequence 
\begin{equation}
\L(j)=\K(j)/M_1 \rightarrow M_1/M_2 \rightarrow M_2/M_3 \rightarrow \cdots \rightarrow M_{n-1}/M_n \ .
\end{equation}
We shall write these composition series vertically, with the head of $\K(j)$ appearing in the first line,
$M_1/M_2$ in the second, {\it etc}. The representation that appears in the last line of the composition series
will be called the {\em socle}. It is the intersection of all (essential)\footnote{A submodule $U$ is essential if 
$U \cap V = 0$ implies $V=0$ for all submodules $V$. In the cases of interest to us, this will always be the 
case.} submodules.  The composition series for the atypical Kac modules are shown in 
Fig.~\ref{Kac_atyp}. Note that for the case of the atypical Kac modules $\K(j)$, both the head and the socle
are isomorphic to the irreducible representation $\L(j)$.  Finally, the composition series of the dual 
Kac module $\K^\vee(j)$ only differs by inverting the grading.\footnote{Note that the irreducible representations
are self-dual, {\it i.e.} $\L_g^\vee(j) = \L_{g-2}(j)$.}

We should stress that the Kac module (or dual Kac module) for 
$j=0$ is special in the sense that the trivial one-dimensional representation $\1 = (0,0)$ appears in its 
composition series. It is important to note that this irreducible representation has grade $-2$,
even though compared to the structure of the Kac module for the other values of $j$, one could have guessed 
that it  has grade $-1$. The operator of grade zero that maps ${\bf 1}_{-2}$ to $\L_{-2}(0)$ is simply
$J^{-}$. 

\begin{figure}[htb] 
\begin{center}
\begin{tikzpicture}

\node (name) at (-4,1) {$\K_0(j \geq \frac{1}{2}):$};

\node (top) at (0,2) {$\L_0(j)$};
\node (left) at (-1.5,1) {$\L_{-1}(j-\frac{1}{2})$};
\node (right) at (1.5,1) {$\L_{-1}(j+\frac{1}{2})$};
\node (bottom) at (0,0) {$\L_{-2}(j)$};

\draw (top) to (left);
\draw (top) to (right);
\draw (bottom) to (left);
\draw (bottom) to (right);

\node (name) at (5,1) {$\K_0(0):$};

\node (top) at (8,2) {$\L_0(0)$};
\node (left) at (7,1) {${\bf 1}_{-2}$};
\node (right) at (9,1) {$\L_{-1}(\frac{1}{2})$};
\node (bottom) at (8,0) {$\L_{-2}(0)$};

\draw (top) to (left);
\draw (top) to (right);
\draw (bottom) to (left);
\draw (bottom) to (right);
\end{tikzpicture}
\end{center}

\mycaption{Compostion series of Kac modules. The representation ${\bf 1}$ appearing in $\K(0)$ is the 
trivial, {\it i.e.}\ the one-dimensional, representation of $\psl$.}  \label{Kac_atyp}

\end{figure}

\subsection{Projective Covers} \label{projective covers}

For the construction of the space of states of the underlying conformal field theory another class
of representations, the projective covers, 
play an important role. The projective cover $\P(j_1,j_2)$ of the irreducible representation 
$\L(j_1,j_2)$ is in some sense the 
largest indecomposable $\g$-representation that has $\L(j_1,j_2)$ as its head. More precisely, the 
condition of $\P$ to be {\em projective} means
that for any surjective homomorphism ${\cal A} \twoheadrightarrow {\cal B}$ and any homomorphism 
$\pi: \P \rightarrow {\cal B}$, there exists a homomorphism $\P \rightarrow {\cal A}$ such that the
diagram
\begin{equation}\label{projective}
 \xymatrix{ 
 & \P \ar@{.>}[dl]\ar[d] \\
 {\cal A} \ar@{>>}[r] & {\cal B} }
\end{equation}
commutes. A representation $\P$ is the \textit{projective cover} of ${\cal B}$ if it is 
projective, and if there exists a surjective homomorphism $\pi:\P \rightarrow {\cal B}$ such that no proper 
subrepresentation of $\P$ is mapped onto ${\cal B}$ by $\pi$.\footnote{A surjective homomorphism $\pi$ 
with this property is sometimes also called $\textit{essential}$. For more details on the use of projective 
modules and covers in representation theory see \cite{Humphreys:2008}.} In our context we are interested 
in the atypical case, {\it i.e.}\ ${\cal B}=\L(j)$ --- for the typical case, where $\K(j_1,j_2) = \L(j_1,j_2)$, the 
projective cover is simply $\P(j_1,j_2) = \L(j_1,j_2)$. Any representation
${\cal M}$ with head $\L(j)$ can be mapped onto $\L(j)$, and the projectivity property 
for $\P(j)$ then implies that for any such ${\cal M}$ we have a surjection
$\P(j) \twoheadrightarrow {\cal M}$. Thus 
the projective cover $\P(j)$ is characterised by the property that any representation $\M$
`headed' by $\L(j)$ can be obtained by taking a suitable quotient of $\P(j)$ with respect to a subrepresentation.
Note that this last condition depends on which category of representations $\M$ we consider. In this paper
we will only work with representations that are completely decomposable under the action of $\g^{(0)}$.
This condition excludes, in particular, the Kac module $\K(0)$, since the arrow between 
${\bf 1}_{g-2}$ and $\L_{g-2}(0)$ is induced by $J^-$. 

The projective cover of an irreducible $\L(j)$ can be constructed by using a generalised 
BGG duality \cite{Zou:1996, Germoni:1998}, which basically states that the multiplicity of the Kac module 
$\K(j')$ in the Kac composition series\footnote{For the Kac composition series we successively
look for submodules such that $M_j/M_{j+1}$ is a direct sum of Kac modules (rather than a direct sum
of irreducible modules).} of $\P(j)$ equals the multiplicity of the irreducible representation $\L(j')$ in 
the composition series of $\K(j)$. However, two complications arise. First, the generalised BGG duality 
only holds in situations where the multiplicities with which $\L(j)$ appears in $\K(j)$ is trivial.
This problem was solved in \cite{Zou:1996,Troost:2011fd} by lifting 
$\psl$ to $\mathfrak{gl}(2|2)$, thereby making $g$ an additional quantum number.
Then the two copies of $\L(j)$ in $\K(j)$ can be distinguished. 
Additionally, the generalised BGG duality has only 
be shown for finite-dimensional modules so far. In this paper, however, we shall assume that it also holds
in the infinite-dimensional case, at least as long as $j$ is sufficiently large ($j \geq 1$). This assumption 
will, {\it a posterori}, be confirmed by the fact that our analysis leads to sensible results. On the other
hand, for $j\leq \tfrac{1}{2}$, we cannot directly apply BGG duality since $\K(0)$ is not part of our category. 
The projective covers for $j \leq \frac{1}{2}$  will be constructed in 
Section.~\ref{sect_proj_cov_small_j}, using directly the universal property of projective covers described 
above.

Applying the BGG duality to the projective covers of $\P(j)$ with $j\geq 1$, and observing that
$\g_{-1}$ generates the states within a Kac module (so that the arrows between different Kac modules
must come from $\g_{+1}$), we obtain from Fig.~\ref{Kac_atyp} (compare \cite{Gotz:2006qp})
\begin{equation}
\P_g(j): \qquad \K_g(0) \rightarrow \K_{g+1}(j-\tfrac{1}{2}) \oplus \K_{g+1}(j+\tfrac{1}{2}) \rightarrow \K_{g+2}(j) 
\,, \quad j \geq 1 \ ,
\end{equation}
where $g$ denotes again the $\Z$-grading introduced before, with the head of $\P_g(j)$ having
grade $g$. In terms of the decomposition into irreducibe representations we then find (again using
Fig.~\ref{Kac_atyp}) the structure described in Fig.~\ref{Proj_irr}. Note that the projective
cover $\P(j)$ covers both the Kac module $\K(j)$, as well as the dual Kac module $\K^\vee(j)$, since
both of them are headed by the irreducible representation $\L(j)$. 

\begin{figure}[htb] %projective cover for generic j
\begin{center}
\begin{tikzpicture}[scale=0.8]

% Kac modul 1

\node (K1_u) at (12,12) {$\L_0(j)$};
\node (K1_l) at (10,10) {$\L_{-1}(j-\tfrac{1}{2})$};
\node (K1_r) at (14,10) {$\L_{-1}(j+\tfrac{1}{2})$};
\node (K1_d) at (8,8) {$\L_{-2}(j)$};

\draw (K1_u) to (K1_l);
\draw (K1_u) to (K1_r);
\draw (K1_d) to (K1_l);
\draw (K1_d) to (K1_r);

% Kac modul 2

\node (K2_u) at (6,10) {$\L_{+1}(j-\tfrac{1}{2})$};
\node (K2_l) at (4,8) {${\L_0(j-1)}$};
\node (Km) at (12,8) {$2\, \L_0(j)$};
\node (K2_d) at (6,6) {$\L_{-1}(j-\tfrac{1}{2})$};

\draw (K2_u) to (K2_l);
\draw (K2_u) to (Km);
\draw (K2_d) to (K2_l);
\draw (K2_d) to (Km);

% Kac modul 3

\node (K3_u) at (18,10) {$\L_{+1}(j+\tfrac{1}{2})$};
\node (K3_r) at (20,8) {$\L_0(j+1)$};
\node (K3_d) at (18,6) {$\L_{-1}(j+\tfrac{1}{2})$};

\draw (K3_u) to (Km);
\draw (K3_u) to (K3_r);
\draw (K3_d) to (Km);
\draw (K3_d) to (K3_r);

% Kac modul 4

\node (K4_u) at (16,8) {$\L_{+2}(j)$};
\node (K4_l) at (10,6) {$\L_{+1}(j-\tfrac{1}{2})$};
\node (K4_r) at (14,6) {$\L_{+1}(j+\tfrac{1}{2})$};
\node (K4_d) at (12,4) {$\L_0(j)$};

\draw (K4_u) to (K4_l);
\draw (K4_u) to (K4_r);
\draw (K4_d) to (K4_l);
\draw (K4_d) to (K4_r);

% Connections

\draw[dashed] (K1_u) to (K2_u);
\draw[dashed] (K1_l) to (K2_l);
\draw[dashed] (K1_r) to (Km);
\draw[dashed] (K1_d) to (K2_d);

\draw[dashed] (K1_u) to (K3_u);
\draw[dashed] (K1_l) to (Km);
\draw[dashed] (K1_r) to (K3_r);
\draw[dashed] (K1_d) to (K3_d);

\draw[dashed] (K4_u) to (K2_u);
\draw[dashed] (K4_l) to (K2_l);
\draw[dashed] (K4_r) to (Km);
\draw[dashed] (K4_d) to (K2_d);

\draw[dashed] (K4_u) to (K3_u);
\draw[dashed] (K4_l) to (Km);
\draw[dashed] (K4_r) to (K3_r);
\draw[dashed] (K4_d) to (K3_d);

\end{tikzpicture}
\end{center}

\mycaption{The projective cover $\P_0(j)$ for $j \geq 1$ in terms of irreducible components. Solid lines 
correspond to mappings decreasing the grading by $1$, while dashed lines increase it
by $1$. Note that the $\Z$-grading lifts almost the entire degeneracy except for the middle component 
$\L(j)$ with multiplicity $2$.} \label{Proj_irr} 
\end{figure}

\subsubsection{Homomorphisms} \label{canonical_mappings}

Before we come to discuss the projective covers for small $j$, let us briefly describe 
the various  homormorphisms between different projective covers. In some sense the 
`basic' homomorphisms (from which all other homomorphisms can be constructed
by composition) are the homomorphisms (with $\sigma=\pm 1$)
\begin{equation}\label{spm}
s_\sigma^\pm : \P(j) \rightarrow \P(j+\tfrac{\sigma}{2}) \ , 
\end{equation}
where the superscript $\pm$ indicates to which of the two irreducible representations
$\L(j+\tfrac{\sigma}{2})$ the head of $\P(j)$ is mapped to, see Fig.~\ref{dia_M1} for an illustration
of the map $s_{+1}^+$.  We shall denote the image of this map by $\M_\sigma^\pm(j)$, 
\begin{equation}\label{Mdef}
\M_\sigma^\pm(j) \equiv s_\sigma^\pm \bigl(\P(j) \bigr) \ .
\end{equation}
Note that it follows from Fig.~\ref{dia_M1} that the kernel of $s_\sigma^\pm$ is 
isomorphic to $\M_\sigma^\pm(j-\tfrac{\sigma}{2})$. Thus we have the exact sequence
\begin{equation}
0 \longrightarrow \M^\pm_\sigma(j-\tfrac{\sigma}{2}) \overset{{\iota}}{\longrightarrow} \P(j) 
\overset{s^\pm_\sigma}{\longrightarrow}  \M^\pm_\sigma(j) \longrightarrow 0 \ ,
\end{equation}
where ${\iota}$ denotes the inclusion $\M^\pm_\sigma(j-\tfrac{\sigma}{2})\hookrightarrow \P(j)$. 
 
\begin{figure}[tb]   %Visualisation of the action of $s^+_{+1}$
\begin{center}

\begin{tikzpicture}[scale=0.85]

%preimage

%Kac modul 1

\node (K1_u_1) at (4,12) {$\L(j)$};
\node (K1_l_1) at (3,10) {$\L(j-\tfrac{1}{2})$};
\node (K1_r_1) at (5,10) {$\L(j+\tfrac{1}{2})$};
\node (K1_d_1) at (2,8) {$\L(j)$};

\draw[very thick] (K1_u_1) to (K1_l_1);
\draw[very thick] (K1_u_1) to (K1_r_1);
\draw[very thick] (K1_d_1) to (K1_l_1);
\draw[very thick] (K1_d_1) to (K1_r_1);

% Kac modul 2

\node[gray] (K2_u_1) at (1,10) {$\L(j-\tfrac{1}{2})$};
\node[gray] (K2_l_1) at (0,8) {${\L(j-1)}$};
\node (Km_1) at (4,8) {$\textcolor{gray}{2}\, \L(j)$};
\node[gray] (K2_d_1) at (1,6) {$\L(j-\tfrac{1}{2})$};

\draw[gray] (K2_u_1) to (K2_l_1);
\draw[gray] (K2_u_1) to (Km_1);
\draw[gray] (K2_d_1) to (K2_l_1);
\draw[gray] (K2_d_1) to (Km_1);

% Kac modul 3

\node (K3_u_1) at (7,10) {$\L(j+\tfrac{1}{2})$};
\node (K3_r_1) at (8,8) {$\L(j+1)$};
\node (K3_d_1) at (7,6) {$\L(j+\tfrac{1}{2})$};

\draw[very thick] (K3_u_1) to (Km_1);
\draw[very thick] (K3_u_1) to (K3_r_1);
\draw[very thick] (K3_d_1) to (Km_1);
\draw[very thick] (K3_d_1) to (K3_r_1);

% Kac modul 4

\node[gray] (K4_u_1) at (6,8) {$\L(j)$};
\node[gray] (K4_l_1) at (3,6) {$\L(j-\tfrac{1}{2})$};
\node[gray] (K4_r_1) at (5,6) {$\L(j+\tfrac{1}{2})$};
\node[gray] (K4_d_1) at (4,4) {$\L(j)$};

\draw[gray] (K4_u_1) to (K4_l_1);
\draw[gray] (K4_u_1) to (K4_r_1);
\draw[gray] (K4_d_1) to (K4_l_1);
\draw[gray] (K4_d_1) to (K4_r_1);

% Connections

\draw[dashed,gray] (K1_u_1) to (K2_u_1);
\draw[dashed,gray] (K1_l_1) to (K2_l_1);
\draw[dashed,very thick] (K1_r_1) to (Km_1);
\draw[dashed,gray] (K1_d_1) to (K2_d_1);

\draw[dashed,very thick] (K1_u_1) to (K3_u_1);
\draw[dashed,very thick] (K1_l_1) to (Km_1);
\draw[dashed,very thick] (K1_r_1) to (K3_r_1);
\draw[dashed,very thick] (K1_d_1) to (K3_d_1);

\draw[dashed,gray] (K4_u_1) to (K2_u_1);
\draw[dashed,gray] (K4_l_1) to (K2_l_1);
\draw[dashed,gray] (K4_r_1) to (Km_1);
\draw[dashed,gray] (K4_d_1) to (K2_d_1);

\draw[dashed,gray] (K4_u_1) to (K3_u_1);
\draw[dashed,gray] (K4_l_1) to (Km_1);
\draw[dashed,gray] (K4_r_1) to (K3_r_1);
\draw[dashed,gray] (K4_d_1) to (K3_d_1);

%Image

%Kac modul 1

\node[gray] (K1_u_2) at (12,8) {$\L(j+\tfrac{1}{2})$};
\node[gray] (K1_l_2) at (11,6) {$\L(j)$};
\node[gray] (K1_r_2) at (13,6) {$\L(j+1)$};
\node[gray] (K1_d_2) at (10,4) {$\L(j+\tfrac{1}{2})$};

\draw[gray] (K1_u_2) to (K1_l_2);
\draw[gray] (K1_u_2) to (K1_r_2);
\draw[gray] (K1_d_2) to (K1_l_2);
\draw[gray] (K1_d_2) to (K1_r_2);

% Kac modul 2

\node (K2_u_2) at (9,6) {$\L(j)$};
\node (K2_l_2) at (8,4) {${\L(j-\tfrac{1}{2})}$};
\node (Km_2) at (12,4) {$\textcolor{gray}{2}\, \L(j+\tfrac{1}{2})$};
\node (K2_d_2) at (9,2) {$\L(j)$};

\draw[very thick] (K2_u_2) to (K2_l_2);
\draw[very thick] (K2_u_2) to (Km_2);
\draw[very thick] (K2_d_2) to (K2_l_2);
\draw[very thick] (K2_d_2) to (Km_2);

% Kac modul 3

\node[gray] (K3_u_2) at (15,6) {$\L(j+1)$};
\node[gray] (K3_r_2) at (16,4) {$\L(j+\tfrac{3}{2})$};
\node[gray] (K3_d_2) at (15,2) {$\L(j+1)$};

\draw[gray] (K3_u_2) to (Km_2);
\draw[gray] (K3_u_2) to (K3_r_2);
\draw[gray] (K3_d_2) to (Km_2);
\draw[gray] (K3_d_2) to (K3_r_2);

% Kac modul 4

\node (K4_u_2) at (14,4) {$\L(j+\tfrac{1}{2})$};
\node (K4_l_2) at (11,2) {$\L(j)$};
\node (K4_r_2) at (13,2) {$\L(j+1)$};
\node (K4_d_2) at (12,0) {$\L(j+\tfrac{1}{2})$};

\draw[very thick] (K4_u_2) to (K4_l_2);
\draw[very thick] (K4_u_2) to (K4_r_2);
\draw[very thick] (K4_d_2) to (K4_l_2);
\draw[very thick] (K4_d_2) to (K4_r_2);

% Connections

\draw[dashed,gray] (K1_u_2) to (K2_u_2);
\draw[dashed,gray] (K1_l_2) to (K2_l_2);
\draw[dashed,gray] (K1_r_2) to (Km_2);
\draw[dashed,gray] (K1_d_2) to (K2_d_2);

\draw[dashed,gray] (K1_u_2) to (K3_u_2);
\draw[dashed,gray] (K1_l_2) to (Km_2);
\draw[dashed,gray] (K1_r_2) to (K3_r_2);
\draw[dashed,gray] (K1_d_2) to (K3_d_2);

\draw[dashed,very thick] (K4_u_2) to (K2_u_2);
\draw[dashed,very thick] (K4_l_2) to (K2_l_2);
\draw[dashed,very thick] (K4_r_2) to (Km_2);
\draw[dashed,very thick] (K4_d_2) to (K2_d_2);

\draw[dashed,gray] (K4_u_2) to (K3_u_2);
\draw[dashed,very thick] (K4_l_2) to (Km_2);
\draw[dashed,gray] (K4_r_2) to (K3_r_2);
\draw[dashed,gray] (K4_d_2) to (K3_d_2);

\draw[dotted, red,->,very thick] (K1_u_1)  .. controls (10,12) and (10,10) ..  node[right]{$s^+_{+1}$} (K2_u_2);

\end{tikzpicture} 
\end{center}

\mycaption{Illustration of the maps $s^\pm_\sigma: \P(j) \longrightarrow \P(j+\tfrac{\sigma}{2})$ using the 
example of $s^+_{+1}$.} \label{dia_M1}
\end{figure}

\subsubsection{The Projective Covers for $j \leq \frac{1}{2}$} \label{sect_proj_cov_small_j}

The cases of $\P(j)$ with $j=0,\tfrac{1}{2}$ need to be discussed separately, since then
BGG duality would give rise to a Kac composition for $\P(j)$ that contains $\K(0)$; however,
as we have explained before, $\K(0)$ is not completely reducible with respect to $\g^{(0)}$, and
hence should not arise in our category. We therefore have to work from first principles, and 
construct $\P(j)$ by the property that any  representation with head $\L(j)$ has to be covered by 
$\P(j)$.\footnote{Note that the projective covers for $j=0$ and $j=\tfrac{1}{2}$ that were 
suggested in section 2.4.2 of 
 \cite{Gotz:2006qp} do not seem to be consistent with these constraints: 
for their choices of projective covers it is not possible to cover both subrepresentations 
generated from $\L_{\pm 1}(0)$ at the first level of $\P(\tfrac{1}{2})$ by $\P(0)$. Indeed, $\P(\tfrac{1}{2})$
predicts that there is a map from each $\L_{\pm 1}(0)$ to the trivial representation in the middle line of 
$\P(\tfrac{1}{2})$, but according to their $\P(0)$, there is only one arrow from $\L(0)$ to the trivial
representation at the first level, and this arrow cannot cover both maps in $\P(\tfrac{1}{2})$.}

Our strategy to do so is as follows. Since we have already constructed $\P(1)$, we know that the 
subrepresentations of $\P(1)$ are part of our category. In particular, this is the case for the two
subrepresentations  whose head is $\L(\tfrac{1}{2})$ at the first level (and that we shall call
${\cal M}^\pm_{+1}(\tfrac{1}{2})$ by analogy to the above). The condition that both of them 
have to be covered by $\P(\tfrac{1}{2})$ puts then strong constraints on the structure of $\P(\tfrac{1}{2})$. 
Assuming in addition that the projective covers are all self-dual then also fixes the lower
part of the $\P(\tfrac{1}{2})$, and we arrive at the representation depicted in
Fig.~\ref{Proj_1/2}. Note that this just differs from the naive extrapolation of Fig.~\ref{Proj_irr}  by the 
fact that the left most irreducible component in the middle line is missing.

\begin{figure}[htbp]   %The projective covers for j = 0 and j = 1/2
\centering

\subfigure[The projective cover $\P(\tfrac{1}{2})$ of $\L(\tfrac{1}{2})$.]{
\begin{tikzpicture}[scale=0.7]

% Kac modul 1

\node (K1_u) at (12,12) {$\L_0(\tfrac{1}{2})$};
\node (K1_l) at (11,10) {$\L_{-1}(0)$};
\node (K1_r) at (13,10) {$\L_{-1}(1)$};
\node (K1_d) at (10,8) {$\L_{-2}(\tfrac{1}{2})$};

\draw (K1_u) to (K1_l);
\draw (K1_u) to (K1_r);
\draw (K1_d) to (K1_l);
\draw (K1_d) to (K1_r);

% Kac modul 2

\node (K2_u) at (9,10) {$\L_{+1}(0)$};
\node (Km) at (12,8) {$2\, \L_{0}(\tfrac{1}{2})$};
\node (K2_d) at (9,6) {$\L_{-1}(0)$};

\draw (K2_u) to (Km);
\draw (K2_d) to (Km);

% Kac modul 3

\node (K3_u) at (15,10) {$\L_{+1}(1)$};
\node (K3_r) at (16,8) {$\L_{0}(\tfrac{3}{2})$};
\node (K3_d) at (15,6) {$\L_{-1}(1)$};

\draw (K3_u) to (Km);
\draw (K3_u) to (K3_r);
\draw (K3_d) to (Km);
\draw (K3_d) to (K3_r);

% Kac modul 4

\node (K4_u) at (14,8) {$\L_{+2}(\tfrac{1}{2})$};
\node (K4_l) at (11,6) {$\L_{+1}(0)$};
\node (K4_r) at (13,6) {$\L_{+1}(1)$};
\node (K4_d) at (12,4) {$\L_{0}(\tfrac{1}{2})$};

\draw (K4_u) to (K4_l);
\draw (K4_u) to (K4_r);
\draw (K4_d) to (K4_l);
\draw (K4_d) to (K4_r);

% Connections

\draw[dashed] (K1_u) to (K2_u);
\draw[dashed] (K1_r) to (Km);
\draw[dashed] (K1_d) to (K2_d);

\draw[dashed] (K1_u) to (K3_u);
\draw[dashed] (K1_l) to (Km);
\draw[dashed] (K1_r) to (K3_r);
\draw[dashed] (K1_d) to (K3_d);

\draw[dashed] (K4_u) to (K2_u);
\draw[dashed] (K4_r) to (Km);
\draw[dashed] (K4_d) to (K2_d);

\draw[dashed] (K4_u) to (K3_u);
\draw[dashed] (K4_l) to (Km);
\draw[dashed] (K4_r) to (K3_r);
\draw[dashed] (K4_d) to (K3_d);

\end{tikzpicture}
\label{Proj_1/2}
} \qquad
\subfigure[The projective cover $\P(0)$ of $\L(0)$.]{

\begin{tikzpicture}[scale=0.7]

% Kac modul 1

\node (K1_u) at (12,12) {$\L_0(0)$};
\node (K1_r) at (13,10) {$\L_{-1}(\tfrac{1}{2})$};
\node (K1_d) at (10,8) {$\L_{-2}(0)$};

\draw (K1_u) to (K1_r);
\draw (K1_d) to (K1_r);

% Kac modul 2

\node (Km) at (12,8) {$\L_0(0)$};

% Kac modul 3

\node (K3_u) at (15,10) {$\L_{+1}(\tfrac{1}{2})$};
\node (K3_r) at (16,8) {$\L_0(1)$};
\node (K3_d) at (15,6) {$\L_{-1}(\tfrac{1}{2})$};

\draw (K3_u) to (Km);
\draw (K3_u) to (K3_r);
\draw (K3_d) to (Km);
\draw (K3_d) to (K3_r);

% Kac modul 4

\node (K4_u) at (14,8) {$\L_{+2}(0)$};
\node (K4_r) at (13,6) {$\L_{+1}(\tfrac{1}{2})$};
\node (K4_d) at (12,4) {$\L_{0}(0)$};

\draw (K4_u) to (K4_r);
\draw (K4_d) to (K4_r);

% Connections

\draw[dashed] (K1_r) to (Km);

\draw[dashed] (K1_u) to (K3_u);
\draw[dashed] (K1_r) to (K3_r);
\draw[dashed] (K1_d) to (K3_d);

\draw[dashed] (K4_r) to (Km);

\draw[dashed] (K4_u) to (K3_u);
\draw[dashed] (K4_r) to (K3_r);
\draw[dashed] (K4_d) to (K3_d);

\end{tikzpicture}
\label{Proj_0}
}

\mycaption{The projective covers $\P(\frac{1}{2})$ and $\P(0)$.} 

\end{figure}

The same strategy can be applied to determine the projective cover $\P(0)$ of $\L(0)$. Now 
$\P(\tfrac{1}{2})$ contains the two subrepresentations generated by $\L(0)$ in the second line, and
$\P(0)$ has to cover both of them. Again, assuming self-duality then leads to the projective
cover depicted in Fig.~\ref{Proj_0}. There is one more subtlety however: in $\P(0)$ it is consistent
to have only one copy of $\L(0)$ at grade zero in the middle line. In order to understand why this
is so, let us review the reason for the multiplicity of $2$ of the corresponding $\L(j)$ representation
for $j\geq \tfrac{1}{2}$. Let us denote the maps leading to and from the relevant $\L(j)$ representation 
in $\P(j)$ (with $j\geq 1$) by $\phi^\pm_{\pm 1}$ and $\bar{\phi}^{\pm}_{\pm 1}$, see 
Fig.~\ref{Proj_middle_part}. It now follows from the fact that $\P(j+\tfrac{1}{2})$ covers the subrepresentations
generated by $\L(j+\tfrac{1}{2})$ that 
\begin{equation}\label{const1}
\bar{\phi}^-_{-1} \circ \phi^-_{-1} = 0 \qquad \hbox{and} \qquad 
\bar{\phi}^+_{-1} \circ \phi^+_{-1} = 0 \ ,
\end{equation}
since $\P(j+\tfrac{1}{2})$ does not contain the representation $\L(j-\tfrac{1}{2})$ at grade $\pm 2$. 
The same argument applied to the two subrepresentations generated by $\L(j+\tfrac{1}{2})$ leads to
\begin{equation}\label{const2}
\bar{\phi}^-_{+1} \circ \phi^-_{+1} = 0 \qquad \hbox{and} \qquad 
\bar{\phi}^+_{+1} \circ \phi^+_{+1} = 0 \ .
\end{equation}

\begin{figure}[htb]  %projective cover for generic j, middle part highlighted
\begin{center}
\begin{tikzpicture}[scale=0.8]

% Kac modul 1 Nodes

\node[gray] (K1_u) at (12,11) {$\L(j)$};
\node (K1_l) at (10,9) {$\L(j-\tfrac{1}{2})$};
\node (K1_r) at (14,9) {$\L(j+\tfrac{1}{2})$};
\node[gray] (K1_d) at (8,6) {$\L(j)$};

% Kac modul 2 Nodes

\node (K2_u) at (6,9) {$\L(j-\tfrac{1}{2})$};
\node[gray] (K2_l) at (4,6) {${\L(j-1)}$};
\node (Km) at (12,6) {$2\, \L(j)$};
\node (K2_d) at (6,3) {$\L(j-\tfrac{1}{2})$};

% Kac modul 3 Nodes

\node (K3_u) at (18,9) {$\L(j+\tfrac{1}{2})$};
\node[gray] (K3_r) at (20,6) {$\L(j+1)$};
\node (K3_d) at (18,3) {$\L(j+\tfrac{1}{2})$};

% Kac modul 4 Nodes

\node[gray] (K4_u) at (16,6) {$\L(j)$};
\node (K4_l) at (10,3) {$\L(j-\tfrac{1}{2})$};
\node (K4_r) at (14,3) {$\L(j+\tfrac{1}{2})$};
\node[gray] (K4_d) at (12,1) {$\L(j)$};

% Connections

\draw[gray] (K1_u) to (K1_l);
\draw[gray] (K1_u) to (K1_r);
\draw[gray] (K1_d) to (K1_l);
\draw[gray] (K1_d) to (K1_r);

\draw[gray] (K2_u) to (K2_l);
\draw[gray] (K2_d) to (K2_l);

\draw[gray] (K3_u) to (K3_r);

\draw[gray] (K3_d) to (K3_r);

\draw[gray] (K4_u) to (K4_l);
\draw[gray] (K4_u) to (K4_r);
\draw[gray] (K4_d) to (K4_l);
\draw[gray] (K4_d) to (K4_r);

\draw[dashed,gray] (K1_u) to (K2_u);
\draw[dashed,gray] (K1_l) to  (K2_l);
\draw[dashed,gray] (K1_d) to (K2_d);

\draw[dashed,gray] (K1_u) to (K3_u);
\draw[dashed,gray] (K1_r) to (K3_r);
\draw[dashed,gray] (K1_d) to (K3_d);

\draw[dashed,gray] (K4_u) to (K2_u);
\draw[dashed,gray] (K4_l) to (K2_l);
\draw[dashed,gray] (K4_d) to (K2_d);

\draw[dashed,gray] (K4_u) to (K3_u);
\draw[dashed,gray] (K4_r) to (K3_r);
\draw[dashed,gray] (K4_d) to (K3_d);

\draw[very thick] (K2_u) to node[draw=white,fill=white]{$\phi^-_{+1}$} (Km);
\draw[very thick] (K2_d) to node[draw=white,fill=white]{$\bar{\phi}^-_{-1}$} (Km);

\draw[very thick] (K3_u) to node[draw=white,fill=white]{$\phi^-_{-1}$} (Km);
\draw[very thick] (K3_d) to node[draw=white,fill=white]{$\bar{\phi}^-_{+1}$}(Km);

\draw[dashed,very thick] (K1_r) to node[draw=white,fill=white]{$\phi^+_{-1}$} (Km);
\draw[dashed,very thick] (K1_l) to node[draw=white,fill=white]{$\phi^+_{+1}$} (Km);

\draw[dashed,very thick] (K4_r) to node[draw=white,fill=white]{$\bar{\phi}^+_{+1}$} (Km);
\draw[dashed,very thick] (K4_l) to node[draw=white,fill=white]{$\bar{\phi}^+_{-1}$} (Km);

\end{tikzpicture}
\end{center}

\mycaption{The maps $\phi^\pm_{\pm 1}$ and $\bar{\phi}^{\pm}_{\pm 1}$ in $\P(j)$
with $j\geq 1$.} \label{Proj_middle_part}
\end{figure}

\noindent 
Now suppose that there was only one $\L(j)$ component at grade zero in the middle line of 
$\P(j)$. Since this one  $\L(j)$ representation is in the image of all four $\phi^\pm_\sigma$, it would 
follow from the above
that it would be annihilated by all four $\bar{\phi}^\pm_\sigma$. Thus the actual $\P(j)$ would not have 
any of the four lines represented by $\bar{\phi}^\pm_\sigma$, and as a consequence would not be
self-dual. On the other hand, if the multiplicity is $2$, there is no contradiction --- and 
indeed multiplicity $2$ is what the BGG duality suggests. 

It is clear from Fig.~\ref{Proj_1/2} that the situation for $\P(\tfrac{1}{2})$ is essentially identical,
but for $j=0$ things are different since  we do not have the analogues of $\phi^{\pm}_{+1}$ and
$\bar{\phi}^{\pm}_{-1}$ any longer, see Fig.~\ref{Proj_0}. Thus the constraints (\ref{const1}) and (\ref{const2}) 
are automatically satisfied, and do not imply that the multiplicity of $\L(0)$ at grade zero in the middle 
line of $\P(0)$ must be bigger than one.

By construction it is now also clear how to extend the definition of $s^\pm_{\sigma}$ in (\ref{spm}) to
$j=\tfrac{1}{2}$ and $j=0$ (where for $j=0$ obviously only $\sigma=+1$ is allowed). Similarly we extend
the definition of $\M_\sigma^\pm(j)$  as in (\ref{Mdef}).

\section{Physical States} \label{physical_states}

Next we want to describe the conformal field theory whose BRST cohomology describes
the physical string states on AdS$_3\times {\rm S}^3$. For the case where we just have pure 
NS-NS flux, this is the WZW model based on the supergoup PSL$(2|2)$ \cite{Berkovits:1999im}.  
Non-linear sigma-models with supergroup targets lead to logarithmic conformal field theories  
\cite{Schomerus:2005bf,Gotz:2006qp,Saleur:2006tf}. We can therefore apply the 
general ideas of  \cite{Gaberdiel:2007jv,Gaberdiel:2009ug} in order to construct their spectrum.
This is best described as a certain quotient space of the tensor products of projective covers, 
see Section~\ref{sec:spectrum}.

Once we have constructed the spectrum we need to define the BRST operator. For the massless 
sector, the BRST operator of \cite{Berkovits:1999im} can be simplified 
\cite{Dolan:1999dc}, and we can identify it with a suitable operator in the universal enveloping 
algebra of $\psl$. 
There is some subtlety about how this BRST operator can be lifted to the direct sum of projective
covers, see Section~\ref{cohomology}, but once this is achieved, it is straightforward to determine its 
cohomology.  We find that the cohomology agrees precisely with the 
supergravity spectrum of  \cite{Deger:1998nm,de Boer:1998ip}, see Section~\ref{sec:physical}. This 
generalises and refines  the recent analysis of \cite{Troost:2011fd}; in particular, we explain in more detail 
how  left- and right-moving degrees of freedom are coupled together, and we are able
to obtain also the correct spectrum for small KK-momenta. (Naively extending the analysis of 
\cite{Troost:2011fd} to small momenta would not have correctly reproduced the expected result.)

\subsection{The Spectrum}\label{sec:spectrum}

The spectrum of the WZW model based on the supergroup PSL$(2|2)$ can be described in terms of 
representations of the affine Lie superalgebra based on $\psl$. As is familiar from the usual WZW models,
affine representations are uniquely characterised by the representations of the zero modes 
that simply form a copy of $\psl$; these zero modes act on the Virasoro highest weight states. In order to describe
the spectrum of the conformal field theory, we therefore only have to explain which combinations of representations 
of the zero modes appear for left- and right-movers. In fact, in this paper we shall only study these 
massless `ground states', and thus the affine generators will not make any appearance. We hope to
analyse the massive spectrum (for which the affine generators will play an important role)
elsewhere.

The structure of the ground states $\H^{(0)}$ should be determined by the harmonic analysis of the 
supergroup. This point of view suggests  \cite{Quella:2007hr} that $\H^{(0)}$ is the quotient of
$\hat\H$ by a subrepresentation $\N$
\begin{equation}\label{specans}
\H^{(0)} = \hat\H / {\cal N} \ , \qquad \hbox{where} \qquad 
\hat\H = \bigoplus_{(j_1,j_2)} \P(j_1,j_2) \otimes \ol{\P(j_1,j_2)} \ ,
\end{equation} 
and the sum runs over all (allowed) irreducible representations $\L(j_1,j_2)$, with $\P(j_1,j_2)$ the 
corresponding projective cover. The relevant quotient should be such that, with respect to the left-moving
action of $\psl$, we can write
\begin{equation}
\H^{(0)} = \bigoplus_{(j_1,j_2)} \P(j_1,j_2) \otimes \ol{\L(j_1,j_2)} \ ,   \label{asym_H0}
\end{equation}
and similarly with respect to the right-moving action.
Furthermore, the analysis of a specific class of logarithmic conformal field theories in
\cite{Gaberdiel:2007jv,Gaberdiel:2009ug} suggests, that the subrepresentation $\N$ has a general
simple form that we shall explain below. This ansatz was obtained in \cite{Gaberdiel:2007jv} 
for the $(1,p)$ triplet models by
studying the constraints the bulk spectrum has to obey in order to be compatible with the analogue
of the identity boundary condition (that had been previously proposed). In \cite{Gaberdiel:2009ug} 
essentially the same ansatz was used in an example where a direct analogue of the identity boundary 
condition does not exist, and again the resulting bulk spectrum was found to satisfy a number of 
non-trivial consistency conditions, thus justifying the ansatz a posteriori. Given the close structural 
similarity between the projective covers of \cite{Gaberdiel:2009ug} and those of the atypical representations 
above, it seems very plausible that the ansatz of \cite{Gaberdiel:2009ug} will also lead to a sensible
bulk spectrum in our context, and as we shall see this expectation is borne out by our results.

In the following we shall only consider the `atypical' part of $\H^{(0)}$, since, using the mass-shell 
condition,  these are the only representations that appear for the massless string states. Actually, it is
only for these sectors that the submodules $\N$ are non-trivial (since for typical $(j_1,j_2)$, the
projective cover  $\P(j_1,j_2)$ agrees with the irreducible representation $\L(j_1,j_2)$, and
hence $\N$ has to be trivial). 

Following \cite{Gaberdiel:2007jv,Gaberdiel:2009ug} we then propose that the 
subspace $\N$ by which we want to divide out $\hat\H$, is spanned by the subrepresentations
\begin{eqnarray}\label{Ndef}
{\cal N}^\pm_\sigma(j) & = & 
 \left(\tp{s^\pm_\sigma}{\id} - {\id}\otimes \overline{(s^{\; \pm}_{\sigma})^\vee}\right) 
 \left({\P(j-\tfrac{\sigma}{2})}\otimes \overline{\P(j)} \right) \ ,
\end{eqnarray}
where $s^\pm_\sigma$ was defined in sect. \ref{canonical_mappings}, and 
$j \geq \max\{0,\tfrac{\sigma}{2}\}$ with $\sigma = \pm 1$. It is easy to see from the 
definition of $s^{\; \pm}_{\sigma}$, see Fig.~\ref{dia_M1},  that the dual homomorphism equals
\begin{equation}
(s^{\; \pm}_{\sigma})^\vee = s^{\; \mp}_{-\sigma} \ .
\end{equation}
Together with (\ref{Mdef}), we can then write the two terms as 
\begin{equation}
\begin{array}{rl}
\tp{s^\pm_\sigma}{\id}\left(\tp{\P(j-\tfrac{\sigma}{2})}{\P(j)} \right) &
= \tp{\M^\pm_\sigma(j-\tfrac{\sigma}{2})}{\P(j)}  \subset \bigl(\P(j) \otimes \overline{\P(j}) \bigr) \ 
\\[0.25cm]
{\id}\otimes \ol{s}^{\; \mp}_{-\sigma} \left(\tp{\P(j-\tfrac{\sigma}{2})}{\P(j)} \right) &
= \tp{\P(j-\tfrac{\sigma}{2})}{\M^\mp_{-\sigma}(j)} \subset 
\bigl(\P(j-\tfrac{\sigma}{2}) \otimes \overline{\P(j-\tfrac{\sigma}{2}})\bigr) \ ,
\end{array} \label{submodules}
\end{equation}
and therefore the two subrepresentations in (\ref{Ndef}) are individual subrepresentations
of different direct summands of $\hat\H$. Dividing out by $\N$ therefore identifies 
\begin{equation}
\bigl(\P(j) \otimes \overline{\P(j}) \bigr)
\supset \ 
\tp{\M^\pm_\sigma(j-\tfrac{\sigma}{2})}{\P(j)} \ \sim \  \tp{\P(j-\tfrac{\sigma}{2})}{\M^\mp_{-\sigma}(j)}  \ 
\subset \bigl(\P(j-\tfrac{\sigma}{2}) \otimes \overline{\P(j-\tfrac{\sigma}{2}})\bigr) \ .
\end{equation}
Note that this equivalence relation does not preserve the $\Z$-grading: for example, 
by considering the corresponding heads, we get the equivalence relation 
\begin{equation}
\bigl(\P(j) \otimes \overline{\P(j}) \bigr)
\supset \ 
\tp {\rep {j-\tfrac{\sigma}{2}}1{\pm1}} {\rep j00} \ \sim \ \tp {\rep {j-\tfrac{\sigma}{2}}00} {\rep j1{\mp 1}}
 \ 
\subset \bigl(\P(j-\tfrac{\sigma}{2}) \otimes \overline{\P(j-\tfrac{\sigma}{2}})\bigr) \ .
\end{equation}
We shall sometimes denote the corresponding equivalence classes by $[\ \cdot \ ]$. It is not difficult
to see that this equivalence relation leads to a description of $\H^{(0)}$ as in eq.~(\ref{asym_H0}). Indeed,
iteratively applying the above equivalence relation we can choose the representative 
in such a way that the right-moving factor, say, is the head of the projective cover; this
is sketched in Fig.~\ref{equi_irreps}.
\begin{figure}[tb]
\begin{center}
\begin{tikzpicture}[scale=0.8, rotate=-45]

\node (P1) at (1,4) {$\tp{\P(j)}{\P(j)}$};

\fill[gray] (0,1) rectangle (2,2);
\draw	(0,0) grid (2,2);
\fill[red] (1,1) circle (4pt);

\node (otimes1) at (2.5,2.5) {$\bigotimes$};

\fill[gray] (3,3) rectangle (5,4);
\draw (2.99,2.99) grid (5,5);
\fill[red] (4,4) circle (4pt);

\node (sim1) at (5.5,5.5) {$\sim$};

\node (P2) at (7,10) {$\tp{\P(j+\frac{\sigma}{2})}{\P(j+\frac{\sigma}{2})}$};

\fill[gray] (6,6) rectangle (8,7);
\draw (5.99,5.99) grid (8,8);
\fill[red] (7,6) circle (4pt);

\node (otimes2) at (8.5,8.5) {$\bigotimes$};

\fill[gray] (9,10) rectangle (11,11);
\draw (8.99,8.99) grid (11,11);
\fill[red] (10,11) circle (4pt);

\node (eq) at (2.5,-3.5) {\huge $=$};

\node (P2) at (7,-2) {$\tp{\P(j+\frac{\sigma}{2})}{\P(j+\frac{\sigma}{2})}$};

\fill[gray] (3,-3) rectangle (4,-1);
\draw (3,-3) grid (5,-1);
\fill[red] (4,-3) circle (4pt);

\node (otimes3) at (5.5,-0.5) {$\bigotimes$};

\fill[gray] (7,0) rectangle (8,2);
\draw (5.99,0) grid (8,2);
\fill[red] (7,2) circle (4pt);

\node (sim2) at (8.5,2.5) {$\sim$};

\node (P3) at (13,4) {$\tp{\P(j+\frac{\sigma+\sigma'}{2})}{\P(j+\frac{\sigma+\sigma'}{2})}$};

\fill[gray] (10,3) rectangle (11,5);
\draw (8.99,2.99) grid (11,5);
\fill[red] (11,3) circle (4pt);

\node (otimes4) at (11.5,5.5) {$\bigotimes$};

\fill[gray] (12,6) rectangle (13,8);
\draw (11.99,5.99) grid (14,8);
\fill[red] (12,8) circle (4pt);
\end{tikzpicture}
\end{center}

\mycaption{Schematic presentation of the equivalence relation. Each big square represents a 
projective cover $\P$, and the shaded regions describe the subrepresentations $\M^\pm_\sigma$ of $\P$. 
The red dots mark exemplary equivalent irreducible components $\tp{\L}{\L}$ in $\tp{\P(j)}{\P(j)}$ and 
$\tp{\P(j+\frac{\sigma}{2})}{\P(j+\frac{\sigma}{2})}$, respectively. Note that by applying the equivalence 
relation, the right-moving irreducible is lifted by one level, while the left-moving one is lowered one level,
until the right-moving irreducible is at the head of some projective cover.} 
\label{equi_irreps}
\end{figure}

Before concluding this subsection, let us briefly comment on possible generalisations 
of our ansatz to WZW models on other supergroups, for example those discussed in 
\cite{Quella:2007hr}\footnote{We thank the referee for suggesting this generalisation to us.}. Let 
us label the irreducible representations by $\lambda$, and their projective covers by $\P(\lambda)$. Thus 
the analogue of (\ref{specans}) is
\begin{equation}\label{specans1}
\H^{(0)} = \hat\H / {\cal N} \ , \qquad \hbox{where} \qquad 
\hat\H = \bigoplus_{\lambda} \P(\lambda) \otimes \ol{\P(\lambda)} \ .
\end{equation} 
In order to construct ${\cal N}$ it is again sufficient to concentrate on the atypical sectors since
otherwise $\P(\lambda)=\L(\lambda)$ is irreducible and the intersection of ${\cal N}$ with
$\P(\lambda) \otimes \ol{\P(\lambda)}$ must be trivial. If $\lambda$ is atypical, on the other hand,
$\P(\lambda)$ is only indecomposable, and it contains a maximal proper submodule that we denote 
by $\M(\lambda)$. Its head is in general a direct sum of irreducible
representations $\L(\mu_i)$. Each direct summand generates a submodule $\M(\mu_i)$ of 
$\P(\lambda)$ which is covered by the projective cover $\P(\mu_i)$. Thus we have the 
homomorphisms $s_{\mu_i} : \P(\mu_i) \rightarrow \P(\lambda)$ via
\begin{equation}
s_{\mu_i}:  \P(\mu_i) \twoheadrightarrow \M(\mu_i) \hookrightarrow \P(\lambda) \ .
\end{equation}
The dual homomorphism  are then of the form 
$s^\vee_{\mu_i}:  \P^\vee(\lambda) \longrightarrow \P^\vee(\mu_i)$, 
where the dual representation $\M^\vee$ is obtained from $\M$ by
exchanging the roles of $\mathfrak{g}_{+1}$ and $\mathfrak{g}_{-1}$. If we assume
the projective covers to be self-dual, $\P^\vee(\mu) = \P(\mu)$, the dual homomorphisms are of the form
\begin{equation}
s^\vee_{\mu_i}: \P(\lambda) \longrightarrow \P(\mu_i) \ .
\end{equation}
It is then again natural to define ${\cal N}$ as the vector space generated by
\begin{equation}
\N_{\mu_i} = \left(\tp{s_{\mu_i}}{\mathrm{id}} - \tp{\mathrm{id}}{s^\vee_{\mu_i}}\right) 
\left(\tp{\P(\mu_i)}{\P(\lambda)}\right) \ .
\end{equation}
By the same arguments as above, the resulting quotient space $\H^{(0)}$  then has the desired form 
\cite{Quella:2007hr}
\begin{equation}
\H^{(0)} = \bigoplus_{\lambda} \tp{\P(\lambda)}{\L(\lambda)} 
\end{equation}
with respect to the left-action. Thus it seems natural that our ansatz for the bulk spectrum
will also apply more generally to WZW models on basic type I supergroups, provided that 
the projective covers are all self-dual $\P^\vee(\lambda) \cong  \P(\lambda)$.

\subsection{The BRST-Operator and its Cohomology} \label{cohomology}

In the hybrid formulation of the superstring every physical state is annihilated by the square of the 
Virasoro zero-mode, $L_0^2\,  \psi = 0$  \cite{Berkovits:1999im}. In this paper we are only
interested in massless physical states. These appear as ground states of affine representations, and 
for them $L_0^2 \, \psi = 0$ is equivalent to the condition that the square 
of the quadratic Casimir  vanishes, $C_2^2\, \psi = 0$. (In fact, it will turn out that in cohomology, 
we will actually have $C_2 \, \psi=0$.) States of this kind appear only
in atypical representations, and hence we can restrict ourselves to the corresponding 
projective covers in $\hat{\cal H}$. Furthermore, because we are only
interested in the ground states, we can ignore the affine excitations.

%The mass-shell condition $L_0=0$ \cite{Berkovits:1999im} implies that massless states are ground 
%states of the affine representation with $C_2=0$. Representations with $C_2=0$ only appear in
%atypical projective covers, and it is thus sufficient to concentrate on these sectors (and ignore the
%affine excitations). 

The cohomological description of the string spectrum \cite{Berkovits:1999im}  
then simplifies, and reduces to the cohomology of the BRST operator
$\Q_\mathrm{hybrid} = K_{ab} S^a_- S^b_-$  \cite{Dolan:1999dc}, as well as its right-moving 
analogue. Here $a$ and $b$ are $\mathfrak{so}(4)$ vector indices, and 
$S^a_- \in \g_{-1}$ while $K_{ab} \in \g^{(0)}$.  Because the $\mathfrak{so}(4)$ indices are all 
contracted, $\Q_\mathrm{hybrid}$ commutes with $\g^{(0)}$, and it follows from a straightforward 
computation that it also commutes with $\g_{-1}$. For the following it will be convenient to define 
more generally
\begin{equation}
\Q_\alpha = K_{ab} S^a_\alpha S^b_\alpha \ , \qquad \alpha = \pm 
\end{equation}
with $\Q_- \equiv \Q_\mathrm{hybrid}$. Note that $\Q_\alpha$ has $\Z$-grading $2 \alpha$. 

{}From now on we shall work with the basis of generators of $\g$ given in Appendix~\ref{app:A}, for which
we have 
\begin{align}
 \Q_\alpha &= -i \bigl[ S^-_{1\alpha}\,S^+_{1\alpha}\,(J^0 + K^0) 
 + S^-_{2\alpha}\,S^+_{2\alpha}\,(J^0 - K^0) \nonumber \\[0.25cm]
 &\phantom{=} \qquad  + S^+_{2\alpha}\,S^-_{1\alpha}\,K^+ 
+ S^-_{2\alpha}\,S^-_{1\alpha}\,J^+ + S^+_{1\alpha}\,S^-_{2\alpha}\,K^- 
+ S^+_{1\alpha}\,S^+_{2\alpha}\,J^-\bigr] \ . 
\end{align}
Using the commutation relations of Appendix~\ref{app:A}, we find by a direct calculation
\begin{equation}
[S^\pm_{m\beta}, \Q_\gamma] = i\, \varepsilon_{\beta\gamma} S^\pm_{m\gamma} \, C_2 \qquad 
\Q_\alpha^2 = S^4_\alpha \,  C_2 \ ,   \label{Q_squared}
\end{equation}
where $S^4_\alpha = S^+_{2\alpha}S^-_{2\alpha}S^+_{1\alpha}S^-_{1\alpha}$, and 
$C_2$ is the quadratic Casimir of $\psl$. Thus if the quadratic Casimir vanishes on a given 
representation ${\cal R}$, $C_2({\cal R})=0$, the operator $\Q_\alpha$ is nilpotent and commutes 
with the full $\psl$ algebra on ${\cal R}$, {\it i.e.}\ it defines a nilpotent $\psl$-homomorphism from 
${\cal R}$ to itself. In particular, the cohomology of $\Q_\alpha$ on ${\cal R}$  then organises itself 
into representations of $\psl$. 

An important class of representations on which the quadratic Casimir vanishes are the 
atypical Kac modules $\K(j)$ with $j\geq \tfrac{1}{2}$. For each $\K(j)$ there are two non-trivial 
homomorphisms $\K(j) \rightarrow \K(j)$: apart from the identity we  have the homomorphism 
$q_-$ that maps the  head of $\K(j)$ to its socle and that has $\Z$-grading $-2$. Since the identity 
operator is not nilpotent, we conclude that the BRST operators $\Q_\pm$ must be equal to 
\begin{equation} 
\hbox{on $\K(j)$:}\qquad \Q_+ = 0 \ , \qquad \Q_-=q_- \ .
\end{equation}
Similarly, on the dual Kac module, $\K^\vee(j)$, the BRST operator $\Q_-$ is trivial, while $\Q_+$ now agrees 
with the non-trivial homomorphism $q_+$ that maps the head of $\K^\vee(j)$ to its socle (which has
now grade $+2$)
\begin{equation} 
\hbox{on $\K^\vee(j)$:}\qquad \Q_+ = q_+ \ , \qquad \Q_-=0 \ .
\end{equation}
\smallskip

\begin{figure}[htb]
\begin{center}
\begin{tikzpicture} [scale = 0.7]
\node (00) at (0,12) {$\rep j00$};

\node (+11) at (6, 9) {$\rep{j+\frac{1}{2}}1{+1}$};
\node (+1-1) at (2, 9) {$\rep{j+\frac{1}{2}}1{-1}$};
\node (-11) at (-2, 9) {$\rep{j-\frac{1}{2}}1{+1}$};
\node (-1-1) at (-6, 9) {$\rep{j-\frac{1}{2}}1{-1}$};

\node[draw,circle] (+20) at (7, 6) {$\rep{j+1}2{0}$};
\node (22) at (3.5, 6) {$\rep{j}2{+2}$};
\node[draw,circle] (20) at (0, 6) {$2\,\rep{j}2{0}$};
\node (2-2) at (-3.5, 6) {$\rep{j}2{-2}$};
\node[draw,circle] (-20) at (-7, 6) {$\rep{j-1}2{0}$};

\node (+31) at (6, 3) {$\rep{j+\frac{1}{2}}3{+1}$};
\node (+3-1) at (2, 3) {$\rep{j+\frac{1}{2}}3{-1}$};
\node (-31) at (-2, 3) {$\rep{j-\frac{1}{2}}3{+1}$};
\node (-3-1) at (-6, 3) {$\rep{j-\frac{1}{2}}3{-1}$};

\node (40) at (0,0) {$\rep j40$};

\draw (00) to  (+11);
\draw (00) to  (+1-1);
\draw (00) to  (-11);
\draw (00) to  (-1-1);

\draw (+11) to (+20);
\draw (+11) to (22);
\draw (+11) to (20);
\draw (+1-1) to (+20);
\draw (+1-1) to (20);
\draw (+1-1) to (2-2);
\draw (-11) to (-20);
\draw (-11) to (22);
\draw (-11) to (20);
\draw (-1-1) to (-20);
\draw (-1-1) to (2-2);
\draw (-1-1) to (20);

\draw (+20) to (+31);
\draw (22) to (+31);
\draw (20) to (+31);
\draw (+20) to (+3-1);
\draw (20) to (+3-1);
\draw (2-2) to (+3-1);
\draw (-20) to (-31);
\draw (22) to (-31);
\draw (20) to (-31);
\draw (-20) to (-3-1);
\draw (2-2) to (-3-1);
\draw (20) to (-3-1);

\draw (40) to (+31);
\draw (40) to (+3-1);
\draw (40) to (-31);
\draw (40) to (-3-1);

\draw[->, blue,dashed] (00) to [bend left] (22);
\draw[->, blue,dashed] (+1-1) to [bend right] (+31);
\draw[->, blue,dashed] (-1-1) to [bend right] (-31);
\draw[->, blue,dashed] (2-2) to [bend left] (40);

\draw[->, red,dotted,very thick] (00) to [bend right] (2-2);
\draw[->, red,dotted,very thick] (+11) to [bend left] (+3-1);
\draw[->, red,dotted,very thick] (-11) to [bend left] (-3-1);
\draw[->, red,dotted,very thick] (22) to [bend right] (40);
\end{tikzpicture} 
\end{center}

\mycaption{The action of the BRST operators $\qt_+$ (blue, dashed arrows) and
$\qt_-$ (red, dotted arrows) on the projective cover ${\cal P}(j)$ for $j \geq 1$.
The irreducible representations that generate the common cohomology of $\qt_+$ and 
$\qt_-$ have been circled.}\label{Proj_irrep}
\end{figure} 

Next we need to discuss the relation between Kac modules and the full CFT spectrum $\H^{(0)}$. 
Using similar arguments as above, it is not difficult to see that, as a vector space, $\H^{(0)}$ is
isomorphic to
\begin{equation}\label{symm}
\H^{(0)} = \bigoplus_{(j_1,j_2)} \K(j_1,j_2) \otimes \ol{\K(j_1,j_2)} \ .
\end{equation}
On the atypical representations (that correspond to the massless states) the
BRST operators $\Q_\pm$ (defined as acting on the two Kac modules) are then indeed nilpotent. 
However, this definition of $\Q_\pm$ does not agree with the usual zero mode action on $\H^{(0)}$
since (\ref{symm}) is only true as a vector space, but not as a representation of the two superalgebra actions.
(Indeed, with respect to the left-moving superalgebra, say, the correct  action is given by (\ref{asym_H0}).)
In order to define the BRST operators on the full space of states it is therefore more convenient to 
lift $\Q_\pm$ to the projective covers. This requires a little bit of care as the operators 
$\Q_\pm$, as defined above, are not nilpotent 
on $\P(j)$. In fact,  the quadratic Casimir does not vanish on $\P(j)$ since it maps, 
for example, the head of $\P(j)$ to $\L_0(j)$ in the middle line, see Fig.~\ref{Proj_irr}.
However, the projectivity property guarantees that there exist  nilpotent operators
\begin{equation}
\qt_\pm : \P(j) \rightarrow \P(j) \ , \qquad \qt^2_{\pm} = 0  \,,
\qquad  [\psl, \qt_{\pm} ] \bigr|_{\P(j)} = 0 \ .
\end{equation}
For example, for the case of $\Q_-$, we apply (\ref{projective}) with ${\cal A}=\P(j)$ and 
${\cal B}=\K(j)$, and  thus conclude  that there exists a homomorphism 
$\qt_- :\P(j) \rightarrow \P(j)$ such that 
\begin{equation}
\xymatrix{ 
 & \P(j) \ar@{.>}[dl]_{\qt_-} \ar[d]^{\Q_- \, \circ \, \pi_{\K}} &&  \pi_{\K} \circ \qt_- = \Q_- \circ \pi_{\K} \,,  \\
 \P(j) \ar@{->>}[r]^{\pi_{\K}} & \K(j) && }%\ar[r] & 0 }
\end{equation} 
where $\pi_\K$ is the surjective homomorphism from $\P(j)$ to $\K(j)$. 
Furthermore, it follows from the structure of the 
projective cover, see Fig.~\ref{Proj_irr} and
Fig.~\ref{Proj_1/2}, that there is only one homomorphism on $\P(j)$  of $\Z$-grading $-2$, namely the one
that maps the head $\L_0(j)$ of $\P(j)$ to $\L_{-2}(j)$ in the middle line. Its square
vanishes (for example, because there is no homomorphism of $\Z$-grading $-4$), and thus
we conclude that $\qt_-$ is nilpotent. The argument for $\qt_+$ is analogous.
The resulting action of $\qt_-$ and $\qt_+$ on $\P(j)$ with $j\geq 1$ 
is depicted in Fig.~\ref{Proj_irrep}.  For $j=\tfrac{1}{2}$, the analysis is essentially the same,
the only difference being the absence of the left-most irreducible representation in the middle line.

For $j =0$ we can argue along similar lines, however with one small modification. Recall that for $j=0$ 
the Kac module $\K(0)$, see Fig.~\ref{Kac_atyp}, is not part of our category (and a similar statement
applies to the dual Kac module $\K^\vee(0)$). However, our category {\em does} contain an analogue of the 
Kac module for $j=0$, which we shall denote by $\hat\K(0)$. It is the quotient of the projective cover 
$\P(0)$ by the subrepresentation $\M^+_{-1}(\frac{1}{2})$, and likewise for the dual Kac module;
their diagrammatic form is given by
\begin{center}
\begin{tikzpicture}
\node (name) at (0,1) {$\hat\K(0):$};

\node (top) at (2,2) {$\L(0)$};
\node (right) at (3,1) {$\L(\frac{1}{2})$};
\node (bottom) at (2,0) {$\L(0)$};

\draw (top) to (right);
\draw (bottom) to (right);

\node (name) at (6,1) {$\hat\K^\vee(0):$};

\node (top) at (8,2) {$\L(0)$};
\node (right) at (9,1) {$\L(\frac{1}{2})$};
\node (bottom) at (8,0) {$\L(0)$};

\draw[dashed] (top) to (right);
\draw[dashed] (bottom) to (right);
\end{tikzpicture}
\end{center}
The quadratic Casimir vanishes on $\hat{\K}(0)$ and $\hat{\K}^\vee(0)$, and thus
$\Q_\pm$ are nilpotent homomorphisms on $\hat{\K}(0)$ and $\hat{\K}^\vee(0)$.
By the same arguments as above, we can then lift $\Q_\pm$ to nilpotent
homomorphisms $\qt_\pm$  on $\P(0)$, and their structure is given in
Fig.~\ref{Qlift0}. 
\begin{figure}[htbp]   %Action of the BRST Operator on the projective cover for j = 0
\centering
\subfigure[The BRST operator ${Q}_+$ acting on the projective cover $\P(0)$.]{
\begin{tikzpicture}[scale=0.4]

% Kac modul 1

\node (K1_u_1) at (3,12) {$\L(0)$};
\node (K1_r_1) at (4,9) {$\L(\tfrac{1}{2})$};
\node (K1_d_1) at (1,6) {$\L(0)$};

\draw[very thick] (K1_u_1) to (K1_r_1);
\draw[very thick] (K1_d_1) to (K1_r_1);

% Kac modul 2

\node[gray] (Km_1) at (3,6) {$\L(0)$};

% Kac modul 3

\node[gray] (K3_u_1) at (6,9) {$\L(\tfrac{1}{2})$};
\node[gray] (K3_r_1) at (7,6) {$\L(1)$};
\node[gray] (K3_d_1) at (6,3) {$\L(\tfrac{1}{2})$};

\draw[gray] (K3_u_1) to (Km_1);
\draw[gray] (K3_u_1) to (K3_r_1);
\draw[gray] (K3_d_1) to (Km_1);
\draw[gray] (K3_d_1) to (K3_r_1);

% Kac modul 4

\node[gray] (K4_u_1) at (5,6) {$\L(0)$};
\node[gray] (K4_r_1) at (4,3) {$\L(\tfrac{1}{2})$};
\node[gray] (K4_d_1) at (3,0) {$\L(0)$};

\draw[gray] (K4_u_1) to (K4_r_1);
\draw[gray] (K4_d_1) to (K4_r_1);

% Connections

\draw[dashed,gray] (K1_r_1) to (Km_1);

\draw[dashed,gray] (K1_u_1) to (K3_u_1);
\draw[dashed,gray] (K1_r_1) to (K3_r_1);
\draw[dashed,gray] (K1_d_1) to (K3_d_1);

\draw[dashed,gray] (K4_r_1) to (Km_1);

\draw[dashed,gray] (K4_u_1) to (K3_u_1);
\draw[dashed,gray] (K4_r_1) to (K3_r_1);
\draw[dashed,gray] (K4_d_1) to (K3_d_1);

%IMAGE

% Kac modul 1

\node[gray] (K1_u_2) at (12,12) {$\L(0)$};
\node[gray] (K1_r_2) at (13,9) {$\L(\tfrac{1}{2})$};
\node[gray] (K1_d_2) at (10,6) {$\L(0)$};

\draw[gray] (K1_u_2) to (K1_r_2);
\draw[gray] (K1_d_2) to (K1_r_2);

% Kac modul 2

\node[gray] (Km_2) at (12,6) {$\L(0)$};

% Kac modul 3

\node[gray] (K3_u_2) at (15,9) {$\L(\tfrac{1}{2})$};
\node[gray] (K3_r_2) at (16,6) {$\L(1)$};
\node[gray] (K3_d_2) at (15,3) {$\L(\tfrac{1}{2})$};

\draw[gray] (K3_u_2) to (Km_2);
\draw[gray] (K3_u_2) to (K3_r_2);
\draw[gray] (K3_d_2) to (Km_2);
\draw[gray] (K3_d_2) to (K3_r_2);

% Kac modul 4

\node (K4_u_2) at (14,6) {$\L(0)$};
\node (K4_r_2) at (13,3) {$\L(\tfrac{1}{2})$};
\node (K4_d_2) at (12,0) {$\L(0)$};

\draw[very thick] (K4_u_2) to (K4_r_2);
\draw[very thick] (K4_d_2) to (K4_r_2);

% Connections

\draw[dashed, gray] (K1_r_2) to (Km_2);

\draw[dashed, gray] (K1_u_2) to (K3_u_2);
\draw[dashed, gray] (K1_r_2) to (K3_r_2);
\draw[dashed, gray] (K1_d_2) to (K3_d_2);

\draw[dashed, gray] (K4_r_2) to (Km_2);

\draw[dashed, gray] (K4_u_2) to (K3_u_2);
\draw[dashed, gray] (K4_r_2) to (K3_r_2);
\draw[dashed, gray] (K4_d_2) to (K3_d_2);

\draw[dotted, red,->, very thick] (K1_u_1) .. controls (10,10) .. node[above]{$\qt_+$} (K4_u_2);

\end{tikzpicture}
\label{Proj_0_BRST_+2}

} \quad
\subfigure[The BRST operator  $\qt_-$ acting on the projective cover $\P(0)$.]{

\begin{tikzpicture}[scale=0.4]

% Kac modul 1

\node (K1_u_1) at (3,12) {$\footnotesize{\L(0)}$};
\node[gray] (K1_r_1) at (4,9) {$\L(\tfrac{1}{2})$};
\node[gray] (K1_d_1) at (1,6) {$\L(0)$};

\draw[gray] (K1_u_1) to (K1_r_1);
\draw[gray] (K1_d_1) to (K1_r_1);

% Kac modul 2

\node[gray] (Km_1) at (3,6) {$\L(0)$};

% Kac modul 3

\node (K3_u_1) at (6,9) {$\L(\tfrac{1}{2})$};
\node[gray] (K3_r_1) at (7,6) {$\L(1)$};
\node[gray] (K3_d_1) at (6,3) {$\L(\tfrac{1}{2})$};

\draw[gray] (K3_u_1) to (Km_1);
\draw[gray] (K3_u_1) to (K3_r_1);
\draw[gray] (K3_d_1) to (Km_1);
\draw[gray] (K3_d_1) to (K3_r_1);

% Kac modul 4

\node (K4_u_1) at (5,6) {$\L(0)$};
\node[gray] (K4_r_1) at (4,3) {$\L(\tfrac{1}{2})$};
\node[gray] (K4_d_1) at (3,0) {$\L(0)$};

\draw[gray] (K4_u_1) to (K4_r_1);
\draw[gray] (K4_d_1) to (K4_r_1);

% Connections

\draw[dashed,gray] (K1_r_1) to (Km_1);

\draw[dashed,very thick] (K1_u_1) to (K3_u_1);
\draw[dashed,gray] (K1_r_1) to (K3_r_1);
\draw[dashed,gray] (K1_d_1) to (K3_d_1);

\draw[dashed,gray] (K4_r_1) to (Km_1);

\draw[dashed,very thick] (K4_u_1) to (K3_u_1);
\draw[dashed,gray] (K4_r_1) to (K3_r_1);
\draw[dashed,gray] (K4_d_1) to (K3_d_1);

%IMAGE

% Kac modul 1

\node[gray] (K1_u_2) at (12,12) {$\L(0)$};
\node[gray] (K1_r_2) at (13,9) {$\L(\tfrac{1}{2})$};
\node (K1_d_2) at (10,6) {$\L(0)$};

\draw[gray] (K1_u_2) to (K1_r_2);
\draw[gray] (K1_d_2) to (K1_r_2);

% Kac modul 2

\node[gray] (Km_2) at (12,6) {$\L(0)$};

% Kac modul 3

\node[gray] (K3_u_2) at (15,9) {$\L(\tfrac{1}{2})$};
\node[gray] (K3_r_2) at (16,6) {$\L(1)$};
\node (K3_d_2) at (15,3) {$\L(\tfrac{1}{2})$};

\draw[gray] (K3_u_2) to (Km_2);
\draw[gray] (K3_u_2) to (K3_r_2);
\draw[gray] (K3_d_2) to (Km_2);
\draw[gray] (K3_d_2) to (K3_r_2);

% Kac modul 4

\node[gray] (K4_u_2) at (14,6) {$\L(0)$};
\node[gray] (K4_r_2) at (13,3) {$\L(\tfrac{1}{2})$};
\node (K4_d_2) at (12,0) {$\L(0)$};

\draw[gray] (K4_u_2) to (K4_r_2);
\draw[gray] (K4_d_2) to (K4_r_2);

% Connections

\draw[dashed,gray] (K1_r_2) to (Km_2);

\draw[dashed,gray] (K1_u_2) to (K3_u_2);
\draw[dashed,gray] (K1_r_2) to (K3_r_2);
\draw[dashed,very thick] (K1_d_2) to (K3_d_2);

\draw[dashed,gray] (K4_r_2) to (Km_2);

\draw[dashed,gray] (K4_u_2) to (K3_u_2);
\draw[dashed,gray] (K4_r_2) to (K3_r_2);
\draw[dashed,very thick] (K4_d_2) to (K3_d_2);

\draw[dotted, red,->, very thick] (K1_u_1) edge[bend left] node[above]{$\qt_-$} (K1_d_2);

\end{tikzpicture}
\label{Proj_0_BRST_-2}
}

\mycaption{The action of the operators $\qt_\pm$ on $\P(0)$.} \label{Qlift0}

\end{figure}

\subsection{The Physical Spectrum}\label{sec:physical}

According to \cite{Dolan:1999dc}, the (massless) physical states of the string theory are described by
the common cohomology of  $\qt_-$ and $\bar{\qt}_{-}$, where $\bar{\qt}_{\pm}$
are the corresponding right-moving BRST operators. Since $\qt_-$ and 
$\bar{\qt}_{-}$ commute with one another, the common cohomology simply consists
of those states that are {\em simultaneously} annihilated by $\qt_-$ and $\bar{\qt}_{-}$, 
modulo states that are {\em either} in the image of $\qt_-$ or $\bar{\qt}_{-}$.

Given the explicit form of the various BRST operators, see 
Fig.~\ref{Proj_irrep} and Fig.~\ref{Qlift0}, it is clear that on the actual space of states 
(\ref{specans}), we have the equivalences
\begin{equation}\label{equiva}
\qt_\pm \otimes \bar{\rm id} \ \cong \ {\rm id} \otimes \bar{\qt}_{\mp} \ .
\end{equation}
We may therefore equivalently characterise the (massless) physical string states as lying in the
common BRST cohomology of $\qt_-$ and $\qt_+$. Note that since $\qt_-$ and 
$\bar{\qt}_{-}$ obviously commute, the same must be true for $\qt_-$ and $\qt_+$; 
this can be easily verified  from their explicit action on the projective covers. 

Since these two BRST operators now only act on the left-movers, we can work with the
representatives as described in (\ref{asym_H0}). From the description of the BRST operators,
see in particular Fig.~\ref{Proj_irrep}, we conclude that the common cohomology of 
$\qt_\pm$ equals for $j\geq 1$
\begin{equation}
H^{0}\left(\P(j)\right) \simeq \L(j-1) \oplus 2\, \L(j) \oplus \L(j+1)\ , \qquad j \geq 1 \ . 
\end{equation}
For $j=\tfrac{1}{2}$, the only difference is the absence of the left-most irreducible representation
in the middle line, and we have instead 
\begin{equation}
H^{0}\left(\P(\tfrac{1}{2})\right) \simeq   2\, \L(\tfrac{1}{2}) \oplus \L(\tfrac{3}{2})\ ,
\end{equation}
while for $j=0$ we get from Fig.~\ref{Qlift0}
\begin{equation}
H^{0}\left(\P(0)\right) \simeq  \L(0) \oplus \L(1)\ . 
\end{equation}
Here both $\L(0)$ and $\L(1)$ appear in the middle line of $\P(0)$, and
$\L(0)$ is the middle of the three $\L(0)$'s.

The actual cohomology of interest is then simply the tensor product of these BRST 
cohomologies for the left-movers, with the irreducible head coming from the right-movers; thus we get
altogether
\begin{eqnarray}
\H_{\rm phys} & = & \Bigl[\Bigl( \L(0) \oplus \L(1) \Bigr) \otimes \ol{\L(0)} \Bigr] \ \oplus \ 
\Bigl[ \Bigl( 2 \L(\tfrac{1}{2}) \oplus \L(\tfrac{3}{2}) \Bigr) \otimes \ol{\L(\tfrac{1}{2})} \Bigr]\nonumber \\
& & \  \oplus \ 
\bigoplus_{j\geq 1} \ \Bigl[ \Bigl( \L(j-1) \oplus 2 \L(j) \oplus \L(j+1) \Bigr) \otimes \ol{\L(j)} \Bigr] \nonumber \\
& = & \Bigl( \L(0) \otimes \ol{\L(0)} \, \Bigr) \ \oplus \ \Bigl( \L(0) \otimes \ol{\L(1)}\,  \Bigr)  \ \oplus \
\Bigl( \L(1) \otimes \ol{\L(0)} \, \Bigr)  \\
& & \ \oplus \ 
\bigoplus_{j\geq \frac{1}{2}} \Bigl[ 
\Bigl( \L(j+1) \otimes \ol{\L(j)}\, \Bigr) \ \oplus \ 
2 \Bigl( \L(j) \otimes \ol{\L(j)} \, \Bigr) \ \oplus \  
\Bigl( \L(j) \otimes \ol{\L(j+1)} \, \Bigr) \Bigr] \ . \nonumber
\end{eqnarray}
The spectrum for $j \geq 1$ fits directly the KK-spectrum of supergravity on AdS$_3 \times {\rm S}^3$
\cite{Deger:1998nm,de Boer:1998ip}, as was already confirmed in  \cite{Troost:2011fd}. It therefore remains
to check the low-lying states. In order to compare our results with \cite{Deger:1998nm,de Boer:1998ip}, 
we decompose
the physical spectrum with respect to the $\mathfrak{su}(2) \oplus \mathfrak{su}(2)$ Lie algebra\footnote{These
generators span the isometry group $\mathfrak{so}(4)=\mathfrak{su}(2) \oplus \mathfrak{su}(2)$  of 
${\rm S}^3$.} corresponding to the bosonic Lie generators $K^a$ and $\bar{K}^a$; the 
relevant representations are therefore labelled by $(j_2,\bar{\jmath}_2)$. For the first few
values of $(j_2,\bar{\jmath}_2)$, the multiplicities are worked out in Tab.~\ref{so4_decom}.
The multiplicities of the last column reproduce precisely the results of \cite{de Boer:1998ip},
see eq.\ (6.2) of that paper with $n_T=1$.

\begin{table} [htb]
\begin{center}
\begin{tabular}{c|c|c|c|c}
$(j_2,\bar{\jmath}_2)$ & $\psl$-rep & \# in $\psl$-rep & \# in $\H$ & $\sum$ \\
\hline
\hline
\multirow{2}{*}{$(0,0)_{S^3}$} 	& $\tp{\L(0)}{\L(0)}$ 	& 4 & 4 & \multirow{2}{*}{6} \\
						& $\tp{\L(\tfrac{1}{2})}{\L(\tfrac{1}{2})}$ &1 & 2 & \\
\hline
\multirow{3}{*}{$(0,\tfrac{1}{2})_{S^3}$} 	& $\tp{\L(0)}{\L(0)}$ 	& 2 & 2 & \multirow{3}{*}{8} \\
								& $\tp{\L(\tfrac{1}{2})}{\L(\tfrac{1}{2})}$ & 2 & 4 & \\
								& $\tp{\L(0)}{\L(1)}$ 	& 2 & 2 & \\
\hline
\multirow{5}{*}{$(\tfrac{1}{2},\tfrac{1}{2})_{S^3}$} 	& $\tp{\L(0)}{\L(0)}$ 	& 1 & 1 & \multirow{5}{*}{13}\\
										& $\tp{\L(\tfrac{1}{2})}{\L(\tfrac{1}{2})}$ & 4 & 8 & \\
										& $\tp{\L(0)}{\L(1)}$ 	& 1 & 1 & \\
										& $\tp{\L(1)}{\L(0)}$ 	& 1 & 1 & \\
										& $\tp{\L(1)}{\L(1)}$ 	& 1 & 2 & \\
\hline
\multirow{3}{*}{$(0,1)_{S^3}$} 	& $\tp{\L(0)}{\L(1)}$ 	& 4 & 4 & \multirow{3}{*}{7} \\
						& $\tp{\L(\tfrac{1}{2})}{\L(\tfrac{1}{2})}$ & 1 & 2 & \\
						& $\tp{\L(\tfrac{1}{2})}{\L(\tfrac{3}{2})}$ 	& 1 & 1 & \\

\end{tabular}
\end{center}

\mycaption{Decompostion of $\H_{\rm phys}$ under $\mathfrak{so}(4)$. The first column denotes the 
$\mathfrak{so}(4)$ representations, the second enumerates the irreducible $\psl$ representations which 
contain the relevant $\mathfrak{so}(4)$ representation. The third column lists its multiplicity within the 
$\psl$ representation, and the fourth its overall multiplicity in $\H_{\rm phys}$. Finally, the last column 
sums the multiplicities from the different $\psl$ representations.} \label{so4_decom}
\end{table}

\section{Conclusions}\label{concl}

In this paper we have given a detailed description of the PSL$(2|2)$ WZW model that underlies the
hybrid formulation of AdS$_3\times {\rm S}^3$ for pure NS-NS flux. Following recent insights into the 
structure of logarithmic conformal field theories 
\cite{Schomerus:2005bf,Gotz:2006qp,Saleur:2006tf,Gaberdiel:2007jv,Gaberdiel:2009ug,Troost:2011fd}
one expects that the space of states has the structure of a quotient space of a direct sum of 
tensor products of projective covers. In this paper we have worked out the details of this proposal: in 
particular, we have given a fairly explicit description of all the relevant projective covers and explained in
detail how the quotient space can be defined. 

In the hybrid formulation the corresponding string spectrum can then be determined from this CFT 
spectrum as a BRST-cohomology. In this paper we have concentrated on the massless states
for which the two BRST operators of  \cite{Berkovits:1999im,Dolan:1999dc} can be written in terms of 
supergroup generators. While these operators are nilpotent on the tensor
product of Kac modules, they are not actually nilpotent on the 
full LCFT space $\H^{(0)}$. However, as we have explained in Section~\ref{sec:physical}, there is a 
natural lift of these  operators to the projective covers, and hence to $\H^{(0)}$. We have described
the structure of the resulting BRST operators in detail and determined their common cohomology. 
The resulting massless string states reproduce precisely the supergravity prediction of  
\cite{Deger:1998nm,de Boer:1998ip}, including the truncation at small KK momenta. 
\smallskip

It would be interesting to extend the BRST analysis to the massive string states. Our ansatz for 
$\H^{(0)}$ makes a concrete proposal for the full LCFT spectrum, and provided we can identify the
general BRST operators of \cite{Berkovits:1999im} in the supergroup language, it should be straightforward 
to work out the full string spectrum in this manner. It would then be interesting to compare this 
to the known string spectrum in the NS-R formalism (again at the WZW point). At least for 
$k\rightarrow \infty$ it was argued in  \cite{Berkovits:1999im}  that the two descriptions
should be equivalent, but there seems to be some debate whether this will continue to hold
at finite $k$ \cite{Gotz:2006qp}. Furthermore, once the identification between the two 
descriptions is established, one could try to understand, for example, the 
non-renormalisation theorem of \cite{deBoer:2008ss} (that was established using NS-R techniques) in the
manifestly spacetime supersymmetric hybrid formulation. We hope to return to these
questions elsewhere.

\section*{Acknowledgements}

This work was partially supported by the Swiss National Science Foundation.

\appendix

\section{Bases and Commutator Relations of $\mathfrak{psl}(2|2)$} \label{app:A}

The Lie superalgebra $\g = \mathfrak{psl}(2|2)$ can be decomposed as
\begin{equation}
 \g = \g_{+1} \oplus \g^{(0)} \oplus \g_{-1} \,,
\end{equation}
where $\g^{(0)}$ is the bosonic subalgebra and $\g^{(1)} = \g_{+1} \oplus \g_{-1}$ gives
the fermionic generators. The bosonic generators are denoted by $K^{ab}$ with 
$\mathfrak{so}(4)$-indices $a,b$ and the fermionic generators are denoted by
$S^a_\alpha \in \g_{\alpha}$ where $\alpha = \pm$. Hence the index $\alpha$ corresponds
to the $\Z$-grading as explained in sect. \ref{intro_superalg}. For later use, we
also define  $\varepsilon_{\alpha\beta}$ as 
\begin{equation}
 \varepsilon_{+-} = -\varepsilon_{-+} = 1 \,, \qquad \varepsilon_{++} = \varepsilon_{--} = 0 \,.
\end{equation}
In the basis used in \cite{Dolan:1999dc, Troost:2011fd}, the commutation relations read
\begin{align*}
[K^{ab},K^{cd}] &=  i \left( \delta^{ac}K^{bd}-\delta^{bc}K^{ad}-\delta^{ad}K^{bc}+\delta^{bd}K^{ac}\right)  \\[2mm]
[K^{ab},S^c_\gamma] &=  i \left( \delta^{ac} S^b_\gamma - \delta^{bc} S^a_\gamma \right) \\[2mm]
[ S^a_\alpha,S^b_\beta] & =  \tfrac{i}{2}\,\varepsilon_{\alpha\beta}\, \varepsilon^{abcd} K_{cd} \ ,
\end{align*}
where indices are raised and lowered with the invariant $\mathfrak{so}(4)$-metric $\delta^{ab}$. 
An appropriate basis change can be made by defining \cite{Gotz:2006qp}
\begin{align*}
  J^0 &= \tfrac{1}{2}\left(K^{12}+K^{34}\right) \\[2mm]
  K^0 &= \tfrac{1}{2}\left(K^{12}-K^{34}\right)  \\[2mm]
  J^\pm &= \tfrac{1}{2} \left(K^{14}+K^{23}\pm iK^{24}\mp iK^{13}\right)  \\[2mm]
  K^\pm &= \tfrac{1}{2}\left(-K^{14}+K^{23}\mp iK^{24}\mp iK^{13}\right)  \\[2mm]
  S_{1\alpha}^\pm &= S_\alpha^1\pm iS_\alpha^2 \\[2mm]
  S_{2\alpha}^\pm &= S_\alpha^3\pm iS_\alpha^4 \ ,
\end{align*}
for which the commutation relations are explicitly given by
\begin{align*}
   [J^0,J^\pm]&\ =\ \pm\,J^\pm &
   [K^0,K^\pm]&\ =\ \pm\,K^\pm\\[2mm]
   [J^0,S_{1\alpha}^\pm]&\ =\ \pm\frac{1}{2}\,S_{1\alpha}^\pm&
   [J^0,S_{2\alpha}^\pm]&\ =\
\pm\frac{1}{2}\,S_{2\alpha}^\pm\\[2mm]
   [K^0,S_{1\alpha}^\pm]&\ =\ \pm\frac{1}{2}\,S_{1\alpha}^\pm&
   [K^0,S_{2\alpha}^\pm]&\ =\
\mp\frac{1}{2}\,S_{2\alpha}^\pm\\[2mm]
   \{S_{1\alpha}^\pm,S_{2\beta}^\pm\}&\ =\ \mp
2\epsilon_{\alpha\beta}\,J^\pm&
   \{S_{1\alpha}^\pm,S_{2\beta}^\mp\}&\ =\ \pm
2\epsilon_{\alpha\beta}\,K^\pm \\[2mm]
   [J^+,S_{1\alpha}^-]&\ =\ S_{2\alpha}^+&
   [J^+,S_{2\alpha}^-]&\ =\ - S_{1\alpha}^+
%\\[2mm]
\end{align*}
\begin{align*}
 [J^-,S_{1\alpha}^+]&\ =\ - S_{2\alpha}^-&
   [J^-,S_{2\alpha}^+]&\ =\  S_{1\alpha}^-\\[2mm]
   [K^+,S_{1\alpha}^-]&\ =\ S_{2\alpha}^-&
   [K^+,S_{2\alpha}^+]&\ =\ - S_{1\alpha}^+\\[2mm] 
   [K^-,S_{1\alpha}^+]&\ =\ - S_{2\alpha}^+&
   [K^-,S_{2\alpha}^-]&\ =\  S_{1\alpha}^-\\[2mm] 
   [J^+,J^-]&\ =\ 2J^0 &
   [K^+,K^-]&\ =\ 2K^0
   \end{align*}
\begin{align*}
   \{S_{1\alpha}^+,S_{1\beta}^-\}&\ =\ %
2\epsilon_{\alpha\beta}\bigl(J^0-K^0\bigr)\\[2mm]
   \{S_{2\alpha}^+,S_{2\beta}^-\}&\ =\ %
2\epsilon_{\alpha\beta}\bigl(J^0+K^0\bigr)\ .
\end{align*}

\noindent The quadratic Casimir operator is
\begin{equation}
 C_2 = C^\text{bos}_2 + C^\text{fer}_2
\end{equation}
with
\begin{align}
C_2^\text{bos} &= - 2 (J^0)^2 - (J^+ J^- + J^- J^+) + 2 (K^0)^2 + (K^+ K^- + K^- K^+) \\
C_2^\text{fer} &=
\frac{\epsilon^{\alpha\beta}}{2}\sum_{m=1}^{2}\left(S^+_{m\alpha}S^-_{m\beta}+S^-_{m\alpha}S^+_{m\beta}\right) 
= \sum_{m=1}^{2}\left(S^+_{m-}S^-_{m+}+S^-_{m-}S^+_{m+}\right) \   . \label{Casfer}
\end{align}
The operator $C_2^\text{fer}$ is the only bilinear in the fermionic generators that commutes with 
the bosonic subalgebra $\bg$.

\end{document}